\documentclass[]{emulateapj}

\def\etal{et al.}

\def\ie{i.e.}
\def\eg{{e.g.}}
\def\ltsima{$\; \buildrel < \over \sim \;$}
\def\simlt{\lower.5ex\hbox{\ltsima}}
\def\gtsima{$\; \buildrel > \over \sim \;$}
\def\simgt{\lower.5ex\hbox{\gtsima}}
\def\fesc{{$\langle f_{esc}\rangle$}}
\def\eq{equation}
\def\pop3{Population~III}
\def\popII{Population~II}
\def\ion#1#2{{\rm #1\,\sc #2}}
\def\HI{{\ion{H}{i} }}
\def\HII{{\ion{H}{ii} }}
\def\GI{{\ion{He}{i} }}
\def\GII{{\ion{He}{ii} }}

\def\fig{Figure}

\def\tab{Table}

\def\hide#1{}

\received{.......}
\accepted{.......}
\slugcomment{to be submitted to ApJ }

\begin{document}
\tighten
\thispagestyle{empty}

\pagestyle{myheadings}
\markright{DRAFT: \today\hfill}

\def\placefig#1{#1}

\title{THE FATE OF THE FIRST GALAXIES. III.\\
PROPERTIES OF PRIMORDIAL DWARF GALAXIES AND THEIR IMPACT ON THE INTERGALACTIC MEDIUM}
\author{Massimo Ricotti\altaffilmark{1}}
\author{Nickolay Y. Gnedin\altaffilmark{2,3,4}} 
\author{J. Michael Shull\altaffilmark{5}}
 
\altaffiltext{1}{Department of Astronomy, University of Maryland \\
College Park, MD 20742. E--mail: ricotti@astro.umd.edu}
\altaffiltext{2}{Particle Astrophysics Center, Fermi National Accelerator
  Laboratory, Batavia, IL 60510.  E--Mail: gnedin@fnal.gov}   
\altaffiltext{3}{Department of Astronomy and Astrophysics, University
  of Chicago, Chicago, IL 60637}
\altaffiltext{4}{Kavli Institute for Cosmological Physics, University of Chicago, Chicago, IL 60637}
\altaffiltext{5}{CASA, Department of Astrophysical \& Planetary Sciences \\
University of Colorado, 389-UCB, Boulder CO 80309.
E--mail: mshull@casa.colorado.edu}
%  .................................................................
 
\begin{abstract}
In two previous papers, we presented simulations of the first galaxies
in a representative volume of the Universe.  The simulations are
unique because we model feedback-regulated galaxy formation, using
time-dependent, spatially-inhomogeneous radiative transfer coupled to
hydrodynamics.  Here, we study the properties of simulated primordial
dwarf galaxies with masses $\simlt 2\times 10^8$ M$_\odot$ and
investigate their impact on the intergalactic medium.
While many primordial galaxies are dark, about 100--500 per comoving
Mpc$^3$ are luminous but relatively faint. They form preferentially in
chain structures, and have low surface brightness stellar spheroids
extending to $20$\% of the virial radius.  Their interstellar medium
has mean density $n_H\approx 10$--$100~{\rm cm}^{-3}$, metallicity
$Z\sim 0.01$--$0.1$ Z$_\odot$ and can sustain a multi-phase
structure. With large scatter, the mean efficiency of star formation
scales with halo mass, $\langle f_*\rangle\propto M_{\rm dm}^{2}$,
independent of redshift.  Because of feedback, halos smaller than a
critical mass, $M_{crit}(z)$, are devoid of most of their
baryons. More interestingly, we find that dark halos have always a
smaller $M_{crit}(z)$ than luminous ones.  Metal enrichment of the
intergalactic medium is inhomogeneous, with only a 1\%--10\% volume
filling factor of enriched gas with [Z/H]$>-3.0$ and 10\%--50\% with
[Z/H] $>-5.0$.  At $z\approx 10$, the fraction of stars with
metallicity $Z<10^{-3}$ Z$_{\odot}$ is $10^{-6}$ of the total stellar
mass.  Although detections of high-redshift dwarf galaxies with the
{\it James Webb Space Telescope} will be a challenge, studies of their
fossil records in the local Universe are promising because of their
large spatial density.
\end{abstract}

\section{Introduction}
\label{sec:int}

In cold dark matter (CDM) cosmologies, the first galaxies in the
universe are predicted to have been about $10^6$ times smaller than
the Milky Way, with characteristic masses comparable to mass estimates
for the smaller dwarf spheroidal galaxies (dSph) observed around our
Galaxy and Andromeda \citep{Mateo:98, Belokurovetal07}.  The
gravitational potentials of these $10^{6-8}$~M$_\odot$ objects are so weak
that the warm and hot ionized phases of their interstellar medium
(ISM) are weakly bound. As a result, each episode of star formation
may produce powerful outflows that could temporarily inhibit further
star formation.

For the sake of brevity, we hereafter refer to ``dwarf primordial"
(dPri) galaxies to indicate galaxies with virial temperature $T_{\rm
vir} \simlt 20,000$ K (or circular velocity $v_c \simlt 20$ km s$^{-1}$).
In contrast to more massive galaxies, the dark matter (DM) halos of
dPri galaxies are too shallow to contain much photoionized gas with
temperatures 10,000--20,000 K. During their formation, the gas is
heated to temperatures below $10^4$~K, where it is unable to cool
by atomic hydrogen (Ly$\alpha$) line emission.  The mass of these DM
halos is $M_{\rm dm} \simlt 2 \times 10^{8}$ M$_\odot$ at their
typical redshifts of formation (redshift $z \simgt 10$). These
galaxies rely primarily on the formation of molecular hydrogen (H$_2$)
to cool and form stars, because metal cooling is negligible as long as
the gas has almost primordial composition.  This situation changes,
after the metallicity of the intergalactic medium (IGM) 
rises above a critical value, $Z_{\rm crit}$,
which could be as large as 1\% solar \citep{SantoroShull:06} at halo
gas densities $n_H \approx$ 10--100 cm$^{-3}$. As the first few stars
are formed, these requirements no longer hold, since some gas
is heated above 10,000 K and is polluted with heavy elements.

The relevance of understanding the formation of the first galaxies is
not purely academic.  Instead, it is closely connected to many outstanding
questions in cosmology, including: 

\noindent 
(1) The {\it Wilkinson Microwave Anisotropy Probe} (WMAP) satellite 
has detected a polarization signal in the
cosmic microwave background radiation (CMB) that indicates that the
optical depth to Thomson scattering is $\tau_e \simeq 0.09 \pm 0.03$ 
\citep{Spergel:07}. This result may require massive star formation 
at redshift $z > 6$ \citep{HaimanB:06, ShullVenkatesan:07}, 
along with partial ionization by X-rays from intermediate-mass black 
holes (BHs) in the first galaxies
\citep{RicottiO:03,MadauR:03, Venkatesan:01,Oh:00}. 

\noindent
(2) The eventual formation of seed BHs in dPri galaxies can be
important for the assembly of supermassive black holes in
the bulges of galaxies and the nature of ultraluminous X-ray (ULX)
sources observed in nearby galaxies \citep[\eg,][]{Miller:03}.

\noindent
(3) Future radio observations of redshifted 21~cm emission from gas at
$6<z<10$ will probe the ionization and thermal history of the IGM prior 
to reionization \citep[\eg,][]{Madau:97}, when dPri galaxies could be 
the dominant population.

\noindent
(4) The origin of the metals observed in the low-density Ly$\alpha$
forest at redshifts $z \sim 2-5$ is an argument of debate (Simcoe \etal\
2004). One view
invokes nearly uniform IGM pre-enrichment produced by the first stars 
at high-redshift \citep{MadauF:01}.  The other view attributes the 
origin of the observed metal lines to hot, metal-enriched superbubbles 
located around Lyman-break galaxies \citep{Adelberger:03}.  Given the 
difficulties associated with both scenarios, it is important to know 
the amount and volume filling factor of metal-enriched IGM produced by 
dPri galaxies.
 
\noindent 
(5) If a substantial population of dPri galaxies existed at redshift
$z \simeq 10$, we expect that about 10\% of these galaxies may survive 
without further mergers to the present \citep{GnedinK:06}. It is a fascinating
possibility that some of the dSph galaxies observed in the Local
Group could be identified as the few well-preserved fossils of dPri galaxies
\citep{RicottiG:05}.

\noindent 
(6) Finally, a new generation of large telescopes will push
the frontiers of the observable universe to the ages when this
primordial galaxy population is forming. Given the current
uncertainties on their importance or even their existence, it is
crucial to make reliable models to predict what these telescopes might
observe.  In particular, infrared observations with the {\it James
Webb Space Telescope} ({\it JWST}) or the {\it Giant Segmented
Mirror Telescope} (GSMT) should be able to constrain theories for
the formation of the first galaxies.

In order to help understand some of the aforementioned cosmological
problems it is essential to know whether a cosmologically significant
number of dPri galaxies formed and to predict their cosmological
impact.  Despite recent progress, the answer to this question is
controversial, largely because of the uncertain effects of radiative
and dynamical feedback from galaxy formation.  In two previous papers
\citep[][hereafter Papers 1 and 2, respectively]{RicottiGSa:02,
RicottiGSb:02} we described our cosmological simulations of
high-redshift galaxy formation with radiative feedback from star
formation.  Those papers dealt primarily with implementing radiative
transfer in the ionizing continuum and understanding both ``positive
and negative feedback" on the formation and destruction of H$_2$.  We
found significant effects of ``radiative feedback" on the first
galaxies and processes of reionization, from redshifts $z \approx 30
\rightarrow 10$.

Our current study (Paper 3) focuses on simulation results on the
population of dwarf primordial galaxies. This paper is organized as
follows. In \S~\ref{sec:data} we describe the set of cosmological
simulations that we analyze in this work.  In \S~\ref{sec:feedback},
we introduce and discuss the feedback processes that may determine the
frequency of the dPri galaxies.  We analyze the cooling mechanisms and
large scale clustering properties (bias) of galaxies to understand
which physical processes are involved in the self-regulation of star
formation and which is dominant. In \S~\ref{sec:igm} we discuss the
processes that eject metals from the galaxies and their importance for
the enrichment of the IGM.  In \S~\ref{sec:stat_prop} we analyze
statistical properties of dPri galaxies such as their mean stellar and
baryon fraction.  We also study their internal properties: the ISM and
the properties of their stellar and dark halos. In \S~\ref{sec:obs} we
address the observability of primordial galaxies at high redshift with
the {\it JWST}. We also address the prospects for the identification
of their fossil records in the local Universe, noting that this topic
has been investigated in greater detail in separate papers
\citep{RicottiG:05, GnedinK:06, Bovill:08}. We show that most low-mass
galaxies in the Local Group are expected to be either dark or too
faint to be detected\footnote{In the last two years, data mining of
the Sloan Digital Sky Survey has revealed the existence of an
ultra-faint population of dwarf spheroidal galaxies that has the same
properties as predicted by our simulations
\protect{\citep{RicottiGSb:02, RicottiG:05, Bovill:08}}. The
galactocentric distribution around the Milky-Way is in agreement with 
predictions for the fossils of the first galaxies
\protect{\citep{GnedinK:06, BovillR:08}}.}. Therefore, the so called
``missing satellite problem'' \citep{Moore:99} 
for the Milky-Way and Andromeda is not a
fundamental challenge to Cold Dark Matter cosmology. We present a
summary of our results in \S~\ref{sec:sum}.

\section{Simulation data}\label{sec:data}

\def\tabone{
\begin{deluxetable*}{clcccccccl}
\footnotesize
\tablecaption{List of simulations with radiative transfer.\label{tab:one}}
\tablewidth{0pt}
\tablehead{
\colhead{Short} & \colhead{RUN} &\colhead{$N_{box}$} & \colhead{$L_{box}$} & \colhead{Mass
  Res.} & \colhead{Res.} & \colhead{$g_\nu$} & 
\colhead{$\epsilon_{UV}$\fesc} & \colhead{$\epsilon_*$} &
\colhead{Comment} \\
\colhead{Name} & \colhead{} & \colhead{} & \colhead{$h^{-1}$ Mpc} & \colhead{$h^{-1}$ M$_\odot$
} & \colhead{$h^{-1}$ pc} & \colhead{} & \colhead{} & \colhead{} & \colhead{}
}
\startdata
S1 & 256L1p3\tablenotemark{d}   & 256 & 1.0  & $4.93\times 10^3$ & 156 &
III\tablenotemark{a} & $2.5\times 10^{-6}$ & 0.1 & {\tiny high resolution run}\\
 \\
S2 & 128L1p2-2\tablenotemark{d} & 128 & 1.0  & $3.94\times 10^4$ & 488 &
II\tablenotemark{b}  & $1.1\times 10^{-7}$ & 0.05 & {\tiny positive feedback}\\
S3 & 128L1f1      & 128 & 1.0  & $3.94\times 10^4$ & 781 & II  &
$1.6\times 10^{-5}$ &0.2 & {\tiny negative feedback}\\
S4 & 128L1noRAD   & 128 & 1.0  & $3.94\times 10^4$ & 781 & - & 0 & 0.2 &
{\tiny without feedback}\\
\\
S5 & 128L1XR\tablenotemark{d}    & 128 & 1.0  & $3.94\times 10^4$ & 488 & II  &
$1.6\times 10^{-5}$ &0.2 &  {\tiny ``X-ray preionization''}\\
S6 & 128L2BH\tablenotemark{d}    & 128 & 2.0  & $3.15\times 10^5$ & 976 & III  &
$1.6\times 10^{-5}$ & 0.2 & {\tiny ``PopIII $\rightarrow$ black holes''}\\
S7 & 128L2PI\tablenotemark{d}    & 128 & 2.0  & $3.15\times 10^5$ & 976 & III  &
$1.6\times 10^{-5}$ & 0.2 & {\tiny ``PopIII $\rightarrow$ PI SNe''}\\
\enddata 

\tablecomments{Parameter description. {\em Numerical parameters:}
  $N_{box}^3$ is the number of grid cells, $L_{box}$ is the box size
  in comoving h$^{-1}$~Mpc and the resolution is in comoving $h^{-1}$~pc. 
  {\em Physical parameters:} $g_\nu$ is the normalized SED (II = \popII\ and
  III = \pop3), $\epsilon_*$ is the star formation efficiency,
  $\epsilon_{UV}$ is the ratio of energy density of the ionizing
  radiation field to the gas rest-mass energy density converted into
  stars (depends on the IMF), and \fesc\ is the escape fraction of
  ionizing photons from the resolution element. Models 1-4 are from
  Paper 1 and are consistent with reionization at $z_{rei}=6$. Model 5
  is from ROG05 and describe a scenario with early pre-ionization
  by X-rays.  Models 6-7 are from RO04 and describe an early Population~III
  reionization consistent with WMAP-1.}
\tablenotetext{a}{$g_\nu$ is modified assuming $\langle f_{esc}\rangle=0.1$,
   $a_0=N_{HeI}/N_{HI}=0.01$ and $a_1=N_{HeII}/N_{HI}=10$ where
   $N_i$ is the column density of the species/ion $i$ (see Paper~1).} 
\tablenotetext{b}{$g_\nu$ is modified assuming $\langle f_{esc} \rangle=0.01$, 
  $a_0=0.1$, $a_1=10$.} 
\tablenotetext{d}{Secondary ionizations included.}
\end{deluxetable*}
}
\placefig{\tabone}

In this work, we study in greater detail the statistical properties of
dPri galaxies in a set of simulations from Paper~2 that include strong
and weak radiative feedback (runs S1, S2, S3 and S4 in
\tab~\ref{tab:one}).  In addition we analyze three simulations from
\cite{RicottiO:03, RicottiOG:03} (hereafter, RO04 and ROG05) that
include pre-ionization by X-rays (run S5) and early ionization by
Pop~III stars with weak and strong SN feedback (run S6 and S7
respectively).  In all the simulations we use a fast method to solve
three-dimensional radiative transfer of \HI, \GI and \GII ionizing
radiation and follow the non-equilibrium chemistry of neutral and
molecular hydrogen and helium. We simulate a cosmologically
representative volume of the universe with initial conditions drawn
from the concordance flat, cold dark matter cosmology with
cosmological constant ($\Lambda$CDM). We stopped the simulations at
redshift $z \sim 8-10$ because of their small volume.  We included a
phenomenological description of star formation and, in some
simulations, the effects of SN feedback.

Our simulations are tailored to study stellar feedback on the baryon
content and star formation in galaxies with masses $M_{\rm dm} \simlt
10^{9}$ M$_\odot$ (\ie, virial temperatures $T_{\rm vir} \simlt 4 \times
10^4$ K).  In our higher resolution simulation we are able to examine
the properties of single objects in detail, since they are resolved
with 10,000--50,000 DM particles and $\sim 10,000$ stellar
particles.  We also analyze the importance of dPri galaxies for metal
enrichment of the high-$z$ IGM. We attempt to understand in more detail 
the mechanisms that trigger and suppress star formation, and the internal
properties of the galaxies that can help us to distinguish these
objects from more massive galaxies that did form at later times by
Ly$\alpha$ cooling.  \cite{RicottiG:05} further evolved the high-resolution 
simulation, including the effects of reionization, and compared the 
properties of the simulated galaxies at redshift $z \sim 8$ with dSph
galaxies observed in the Local Group.  They found that the properties 
of most dSph galaxies are consistent with being the fossils of 
this first population of galaxies.

\section{Feedback-Regulated Galaxy Formation}
\label{sec:feedback}

Since the earliest works \cite[\eg,][]{CouchmanR:86} it has been
realized that in CDM cosmologies, the first subgalactic structures
form as a consequence of the collapse of rare dark matter density
perturbations with masses of $10^5-10^6$ M$_\odot$ at redshifts $z
\sim 30-40$.  The initial gas cooling must be provided by
collisionally excited H$_2$ rotational and vibrational transitions.
\cite{Tegmark:97} estimated that a minimum H$_2$ abundance of $x_{H_2}
\approx 10^{-4}$ is required to trigger star formation in a dark halo
in less than a Hubble time.  In a dust-free gas, H$_2$ formation is
catalyzed by the H$^-$ ion, that forms as a consequence of the shocks
that partially ionize and heat the gas during the virialization
process. At a given redshift, the mass of the smaller halo that can
form stars is determined by its virial temperature and therefore by
its mass. This analytical result has been confirmed by hydrodynamical
cosmological simulations.  \cite{Abel:02} carried out such
numerical simulations for a selected $10^6$ M$_\odot$ halo, using
adaptive mesh refinement, that resolves the collapse over a large range of
scales. They find that, in this selected halo, only one star with mass
between 10--100 M$_\odot$ will probably form.  \cite{BrommCL:99} have
also found similar results using a variety of initial conditions for
the protogalaxies.  These numerical results confirm longstanding
theoretical arguments that the first stars should be
massive: their characteristic mass reflects the larger Jeans mass in
the inefficiently cooling metal-free gas.  However, the cooling by
trace-metal fine-structure lines depends on the gas density
\citep{SantoroShull:06} and coupling with the cosmic microwave
background.  Thus, the Jeans Mass and ``critical metallicity" are
sensitive to the gas density in the halos.

Because of space constraints, we will not discuss the many papers on
the importance of \pop3\ stars and the first galaxies for reionization
\nocite{VenkatesanT:03}(see Venkatesan, Tumlinson, \& Shull 2003 and
references therein).  The typical mass and initial mass function of
the first stars is still not well constrained, because of the
uncertain role of radiative feedback during the final phases of the
protostellar collapse. Even more uncertain is the impact of the first
stars compared to \popII.  This depends on how many \pop3\ stars can
form and the duration of time until they are outnumbered by normal
\popII\ stars \citep{RicottiOI:03}.

\subsection{Negative feedback}

After the first few stars formed, the Universe becomes difficult to
model. The H$_2$ photodissociating radiation in the Lyman-Werner bands
($11.3-13.6$ eV) emitted by the stars themselves can destroy H$_2$ and
inhibit gas cooling.  In addition, the \HI ionizing radiation
(ultraviolet and X-ray photons) emitted by hot stars, black holes, and SN
remnants may become important or dominant in producing the H$^-$ that
catalyzes H$_2$ formation \citep[\eg,][]{HaimanRL:96, Ferrara:98,
RicottiGS:01, Ahn:07, Whalen:08}.  The formation of H$_2$ from
collisionally ionized gas during virialization is still important, but
it might become a subdominant effect, especially if galaxies are
clustered, as observed today.  The first semianalytic
\citep{HaimanAR:00} and numerical \citep{Machacek:00} studies on the
radiative feedback from the first galaxies included the effects of an
H$_2$ dissociating background produced by hot stars.  In these
models, the formation of dPri galaxies is strongly suppressed by the
H$_2$ photodissociating radiation. As a result, efficient and
widespread star formation in the universe is delayed until the
collapse of more massive DM halos ($M_{\rm dm} \simgt 10^9$ M$_\odot$)
at later times ($z \simgt 15$) formed by Ly$\alpha$ cooling
\citep{OhH:02}. \cite{Yoshida:03} and \cite*{Tassis:03} have also
performed simulations on the collapse of pre-galactic clouds, finding
that for radiation in the H$_2$ Lyman-Werner bands with flux
$J>10^{-23}$ ergs s$^{-1}$ cm$^{-2}$ Hz$^{-1}$ sr$^{-1}$, H$_2$
molecules are rapidly dissociated, rendering the gas cooling
inefficient.  They both find that dwarf-sized dark matter halos
assembled prior to reionization are able to form stars and show large
variations in their gas content because of stellar feedback and
photoionization effects.

\subsection{Positive feedback regions}

Local feedback effects were not included in the aforementioned
simulations. \cite{RicottiGS:01} demonstrated the importance of
``positive feedback", finding that shells of H$_2$ can be
created continuously both in precursors around the Str\"omgren spheres
produced by ionizing sources and, for a bursting mode of star
formation, inside recombining \HII regions. Recent studies have
confirmed the existence and importance of positive feedback regions
\citep{JohnsonB:07, Ahn:07, Whalen:08} in greater detail.

This local positive feedback could be important, but it is
difficult to incorporate into cosmological simulations because the
implementation of spatially inhomogeneous, time-dependent radiative
transfer is computationally expensive and challenging. In Paper 1 we
used a fast method to solve radiative transfer coupled to
hydrodynamics in cosmological simulations of a representative volume
of the Universe.  In Paper~2 we explored the importance of different
feedback processes in producing what appears as a self-regulated star
formation mode on cosmological scales. 

The main parameters that regulate the global star
formation rate at high-redshift are $\langle f_{esc}\rangle$, defined
as the fraction of ionizing radiation that escapes from the simulation
resolution element, and the initial mass function (IMF) of the stars.
The intensity of the H$_2$ dissociating background and the assumed
efficiency of star formation have, surprisingly, only a minor effect
on the self-regulation of star formation. Adopting a Salpeter IMF and
$\langle f_{esc}\rangle\simlt 1$\%, we showed that dPri galaxies may
account for most of the stellar mass at redshift $z \sim 9$.  If the
IMF is top-heavy, or if $\langle f_{esc}\rangle\sim 1$, the global star
formation is reduced but not fully suppressed.  Internal sources of
ionizing photons such as massive stars or quasars produce galactic
winds in dPri galaxies that regulate their star formation rates by
reducing the gas supply. As a consequence, their star formation
history is characterized by several short starburst episodes. The
total fraction of stars produced in primordial galaxies depends on the
intensity of these bursts of star formation.  From the simulations, it
appears that dPri galaxies cannot reionize the low-density regions of
the IGM because the Str\"omgren spheres never reach the overlap
phase. However, they can produce and eject a substantial mass in heavy
elements.

\cite{Machacek:03} studied the effect of a moderate X-ray background
on the formation of the first galaxies, finding that they have a minor
effect on the global star formation rate. They concluded that the
feedback does not completely suppress star formation in low-mass
galaxies, in agreement with Paper~2. However, in disagreement with
this study, they find that depending on the intensity of the
dissociating background, star formation is delayed and less efficient
in smaller mass halos.  The reason for this disagreement may be their
neglect of local feedback such as photoevaporation of the ISM from
stellar winds. \cite{Susa:04} performed simulations of the effect of
reionization on star formation in low-mass galaxies, finding that star
formation in halos that collapse prior to reionization is completely
suppressed after reionization if the halo is small.

\section{Local and global feedback}

In hierarchical models, galaxies form preferentially in groups and 
filaments.  The feedback processes that regulate the cooling of
the gas, and therefore star formation in dPri galaxies can be grouped
in two categories, with processes labeled as internal (I1, I2, I3) 
and external (E1, E2, E3, E4):

{\em Internal feedback} produced by star formation inside each
galaxy: (I1) photoionization by massive stars and mini-quasars is
sufficient to evaporate most of the ISM and temporary halt star
formation; (I2) galactic winds produced by SN explosions may become
important after about 10 Myr; (I3) heavy element self-enrichment
affects the ISM properties and possibly the stellar IMF.

{\em External feedback} can operate on local scales or on cosmological
scales: (E1) the reheating of the IGM produced by ionizing photons (UV
and X-rays) increases the Jeans mass of the IGM preventing gas
collapse inside the smaller mass halos; this is a negative feedback on
cosmological scales; (E2) the radiation backgrounds operate on
cosmological scales: the background in the H$_2$ Lyman-Werner bands
(FUV radiation) is a ``negative feedback'' as it dissociates molecular
hydrogen while the X-ray background, increasing the fractional
ionization of the gas, may promote H$_2$ formation and can be a
``positive feedback''; (E3) feedback from UV ionizing radiation is a
local feedback as it operates on galactic scales. Ionizing radiation
produce what we call ``positive feedback regions''
\citep{RicottiGS:01}. These are regions of enhanced H$_2$ formation
located just ahead of ionization fronts and inside recombining \HII
regions; (E4) contamination of the IGM with heavy elements ejected
from neighboring galaxies can promote gas cooling and galaxy formation
and should be considered as a ``positive feedback''.

\def \capfigaa{({\em Left}). Mean molecular hydrogen abundance,
  $x_{H_2}$, at $z=17.5$ as a function of the mean temperature of the
  sink ``stellar particles'' in the simulation S2 (top panel) and S1
  (bottom panel).  ({\em Right}).  Cooling time compared to the Hubble
  time as a function of the redshift of formation of sink ``stellar
  particles''.  The top panel is the simulation S2 at $z=12.5$ and the
  bottom panel the simulation S1 at $z=12.5$. In all panels the colors
  show the metallicity of the star particle: $Z/Z_\odot < 5 \times
  10^{-4}$ (red); $5 \times 10^{-4} < Z/Z_\odot < 5 \times 10^{-3}$
  (yellow); $5 \times 10^{-3}< Z/Z_\odot < 5 \times 10^{-2}$ (green);
  and $Z/Z_\odot > 5 \times 10^{-2}$ (blue).
}
\placefig{
\begin{figure*}[thp]
\epsscale{1.0}
\plottwo{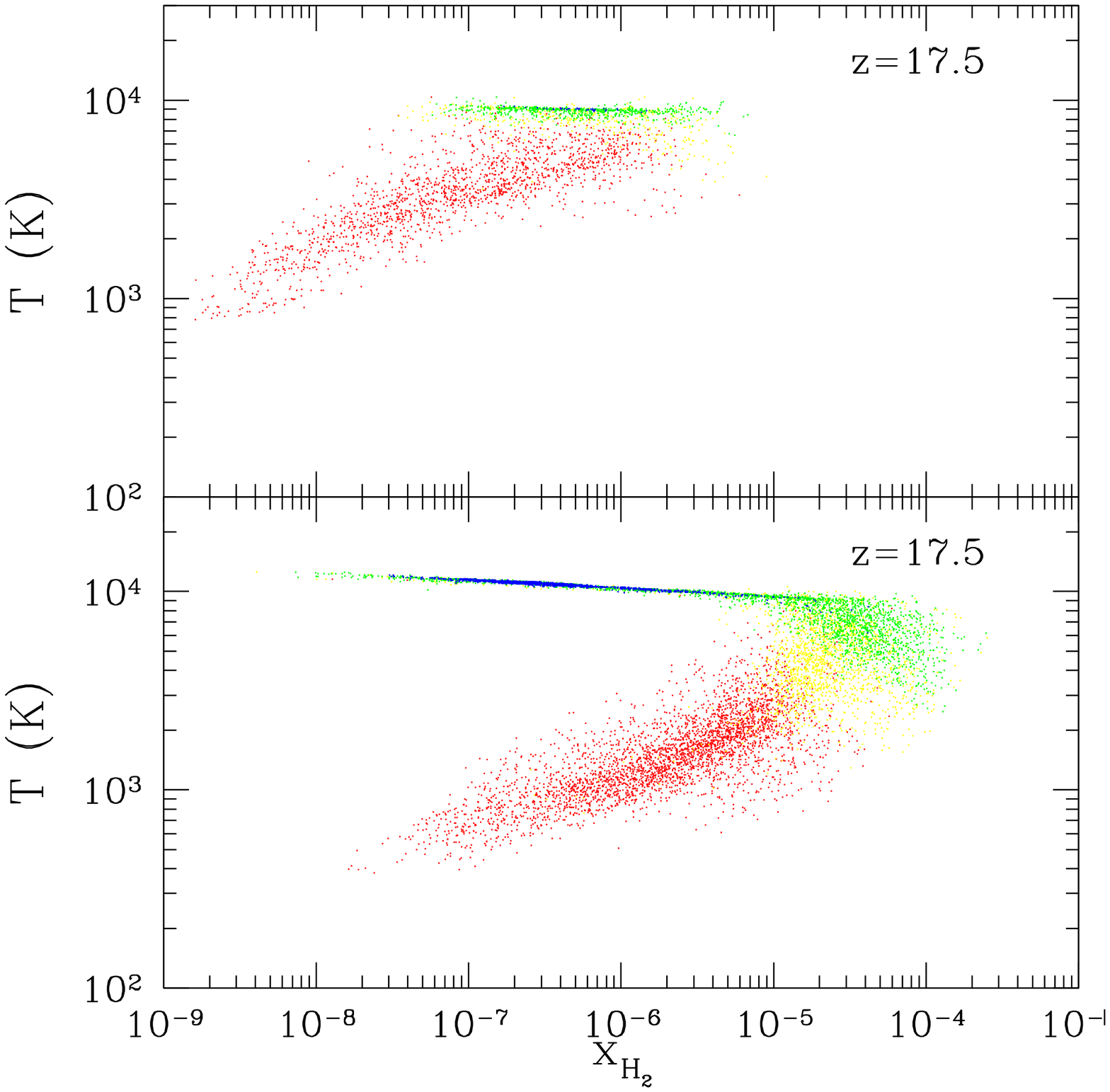}{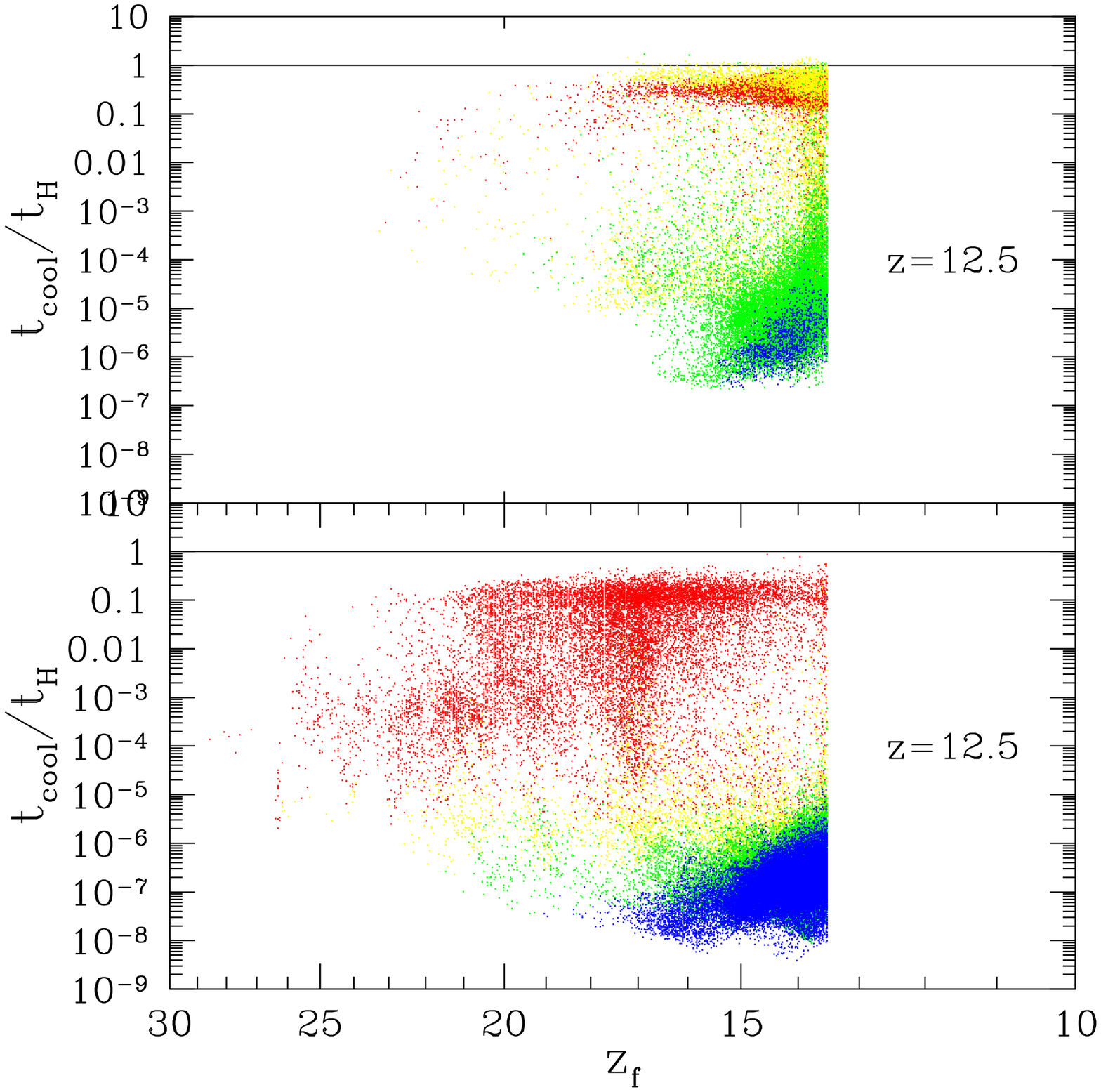}
\caption{\label{fig:tcool}\capfigaa}
\end{figure*}
}

Although these processes play a role in the regulation of stars and
galaxies formation, it is generally possible to identify a dominant
feedback process in each simulation. The dominant feedback is
determined by the value of the free parameters of the simulation such
as $\langle f_{esc}\rangle$, the IMF, and the ionizing spectrum of the
sources. In Paper~2 we found that when the efficiency of emission of
ionizing radiation per baryon converted into stars is large, for 
a top-heavy IMF or Salpeter IMF and $\langle f_{esc}\rangle\sim 1$,
processes (I1) and (E3) are the dominant feedback mechanisms producing
short episodes of bursting star formation in low-mass galaxies. If
instead we assume a Salpeter IMF and $\langle f_{esc}\rangle\ll 1$,
process (E2) dominates and the dissociating radiation background
suppresses the formation of the smaller mass galaxies.
\cite{Venkatesan:01} and \cite{RicottiOG:03} found that, if the first
galaxies host accreting black holes, the IGM is heated to about 10,000~K 
at redshifts $z \sim 20-25$. In this case, process (E1) is 
effective in reducing star formation in low-mass galaxies (see Fig.~2
in ROG05). Finally, with a top-heavy IMF and mechanical feedback from
normal SN or pair-instability SN explosions, process (I2) is important
in suppressing star formation also in galaxies more massive than
$10^8$ M$_\odot$. Metal enrichment (processes [I2] and [E4]) is
important in all simulations (see \S~\ref{ssec:cooling}).

\subsection{Cooling processes}\label{ssec:cooling}

In this section, we focus on the role of different cooling processes
that ultimately lead to the gravitational collapse of the first stars. 
We show that,
after the H$_2$ in the ISM and IGM is destroyed by the dissociating
radiation emitted by the first stars, star formation in a subset of
low-mass halos is not suppressed. This is a consequence of the
positive feedback from the proximity to already formed galaxies or
groups, whose ionizing radiation produces H$^-$ and H$_2$
\citep{RicottiGS:01}.  Only a small fraction of the first stars form
directly from the collapse of low temperature, zero-metallicity
clouds, through cooling by H$_2$ ro-vibrational transitions. The
collapse of most gas clouds is triggered by metal cooling and by Ly$\alpha$
cooling in the photo-heated and metal polluted gas produced by the
first stars.  Zero metallicity stars are efficient catalysts for
further episodes of star formation, not only inside their host halo
but also in nearby halos.  This has important implications for the
clustering and bias properties of the first galaxies, which form
preferentially chain-like structures, analogous to young star clusters
at low redshift.

Metal enrichment from multiple episodes of star formation within a
galaxy (process I2) or metal contamination from neighboring galaxies
(process E4) are important in all simulations.  H$_2$ cooling is
responsible for the collapse of the first proto-star clusters but
subsequently, as the first stars form in a galaxy, their heating and
metal pollution provide the dominant cooling mechanisms (Ly$\alpha$
and metal cooling) and they are seeds for further star formation. In
\fig~\ref{fig:tcool}, we show the temperature versus H$_2$ abundance
of the sink ``stellar particles'' in two simulations.  The figure on
the right shows the cooling time, compared to the Hubble time, as a
function of the redshift of formation of stellar particles.  The top
and bottom panels show simulations S2 and S1, respectively. In both
panels the colors show the metallicity of the star particle (see
caption). Note that in our simulations the star particles lose track
of their initial properties when the star formation is continuous, and
the metallicity and H$_2$ abundance of the star particle is the
mean mass, weighted over time. When star formation in a cell is stopped
by feedback, the star particle is released with its properties. A new
particle will be created in the same cell if it experiences a new
burst of star formation.

The H$_2$ abundance in the collapsing ``stellar particles'' is about
$x_{H_2} \sim 10^{-5} - 10^{-6}$, lower than the value $x_{H_2} \sim
10^{-4}$ derived by \cite{Tegmark:97}. However, the overdensity in the
core of a newly virialized dark halo (\ie, with gas temperature equal
to the virial temperature) is about 100 times larger than the mean
overdensity $n_{\rm vir}$ adopted in their work. This ensures that, as
shown in \fig~\ref{fig:tcool} (right), the cooling time is shorter
than the Hubble time at any given redshift. The reaction $H^- + H
\rightarrow H_2+e^-$ that dominates the formation of H$_2$ absorbs
kinetic energy from the gas, producing a net cooling rate that is also
important at low temperatures.

In summary, the relative importance of cooling and feedback processes
depends on the assumed IMF and $\langle f_{esc}\rangle$.  
Generally, H$_2$ cooling is
important for the formation of the first few stars in each
protogalaxy. After the first episode of star formation, if most gas
has not been blown out, Ly$\alpha$ and metal lines become the dominant
coolants. The strong clustering of the first dark halos also promotes
positive feedback trough metal contamination and photoionization
(that promotes H$_2$ formation) of neighbor galaxies and satellites.

\subsection{Clustering of Primordial Galaxies}\label{subsec:bias}

\def\capfigba{Clustering properties of first luminous galaxies.
  We show positions of dark halos with $M_{dm}>10^6$ M$_\odot$ in the
  simulation S1 projected on the $x-y$ plane in a slice with $\Delta z
  = 0.2~h^{-1}$~Mpc at $z=17.5$, 14.6, 12.5, and 10.2
  (clockwise from top-left panel). Black circles show halo 
  virial radii, and colored symbols mark halos
  hosting a luminous galaxy with $L_V > 5 \times 10^5$ L$_\odot$
  (yellow), $5 \times 10^4 <L_V < 5 \times 10^5$ L$_\odot$ (cyan) and
  $L_V < 5 \times 10^4$ L$_\odot$ (red). We assume
  $M_*/L_V=1/50$ (solar), appropriate for a young stellar population. 
  Most luminous galaxies seem to form in groups or filaments, with few in
  isolation in the lower density IGM.}  \placefig{
 \begin{figure*}[thp]
 \epsscale{1.0}
\plotone{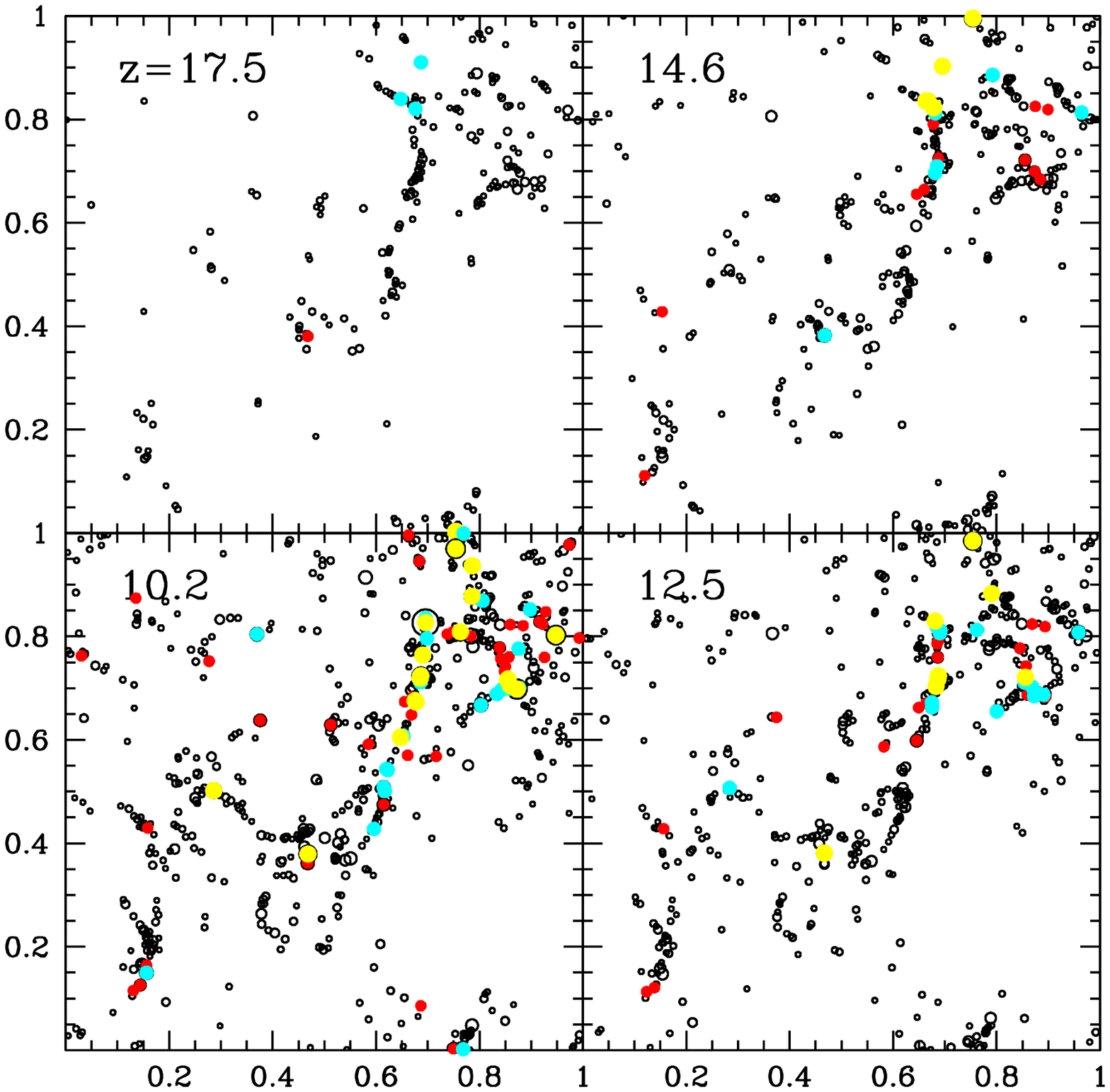}
 \caption{\label{fig:bias1}\capfigba}
 \end{figure*}
} 
\def\capfigba{Same as in \protect{\fig~\ref{fig:bias1}} but showing
the efficiency of star formation efficiency $f_*=M_*/M_{bar}^{max}$,
where $M_{bar}^{max}=M_{dm}\Omega_b/\Omega_m$, rather
than the luminosity of dwarf galaxies. Black circles show dark halos
with mass $>10^6$ M$_\odot$ and yellow dots show luminous halos with
mass $>10^7$ M$_\odot$. The blue and magenta symbols show luminous
halos with mass $<10^7$ M$_\odot$ with $f_*>10^{-3}$ and
$f_*<10^{-3}$, respectively. In addition, luminous halos with a mass
$<3 \times 10^6$ M$_\odot$ are shown as a star rather than a circle.
The plots illustrate the importance of local positive feedback in
halos with $M<10^7$ M$_\odot$: halos with identical masses can be
either dark or luminous, but are likely luminous if they are nearby
another luminous halos.}  \placefig{
\begin{figure*}[thp]
\epsscale{1.0}
\plotone{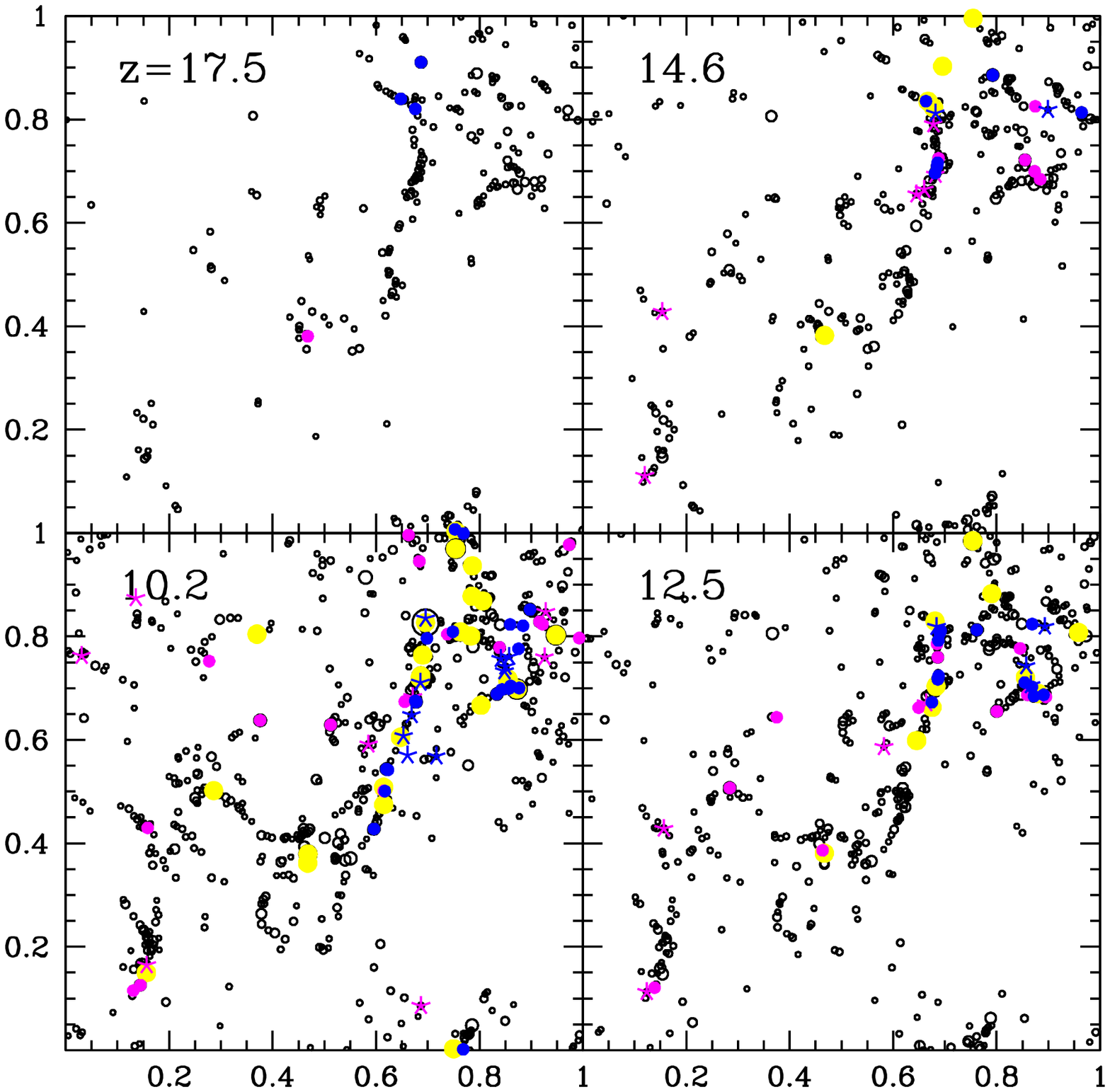}
 \caption{\label{fig:bias2}\capfigba}
 \end{figure*}
}
Most galaxies in the Local Volume are approximately located in a sheet
or filament also known as the ``Supergalactic'' plane.  Presently,
within $5$ Mpc from our Galaxy, only about $2-3$ faint dwarf galaxies
have been discovered that are not associated with any luminous galaxy:
Tucana \citep{Lavery:92}, Cetus \citep{Whiting:99}, and perhaps the
recently discovered Apples1 \citep{Pasquali:05}.  The paucity of dwarf
galaxies out of the Supergalactic plane is a test for models of the
formation of dPri galaxies. Unfortunately, the volume of our
simulations is too small to allow us to study the spatial
distribution of low-mass galaxies at $z=0$. In order to
answer this important question, \cite{BovillR:08} developed a new
method to follow the halo evolution of the first galaxies from the
redshift of formation to $z=0$. Already at redshift $z \sim 10$,
low-mass galaxies are highly biased, reflecting the importance of
local feedback processes in our simulations. Few low-mass galaxies are
observed in isolation, and typically these are the faintest of the
population.

We find that luminous galaxies form preferentially near previously
formed galaxies and are highly biased. This reflects the importance of
local positive feedback (\ie, the H$_2$ precursor in front of \HII
regions and metal enrichment) in promoting star formation.
\fig~\ref{fig:bias1} shows the projected positions of dark halos of
mass $M_{\rm dm}>10^6$ M$_\odot$ in a slice of the run S1 at four
different redshifts.  The sizes of the black circles are proportional
to the virial radii of the dark halos, and the filled circles mark
halos hosting luminous galaxies, color coded according to their
luminosity: red being the faintest and yellow the brightest
galaxies. We have assumed $M_*/L_V =1/50$ (solar units), appropriate
for starbursts at $t \sim 100$ Myr.  It appears that luminous galaxies
are more clustered than dark halos of the same mass, and line up along
the dark matter filaments. Inspecting the four panels in
\fig~\ref{fig:bias1}, it may appear that a wave of star formation
propagates along the dark filaments triggered by the formation of a
first galaxy. This is somewhat analogous to what is observed for star
clusters in the ISM of galaxies.  Although, \fig~\ref{fig:bias1} does
not convincely demonstrate the existence of a propagating wave of star
formation, the importance of local positive feedback can be
demonstrated rigorously. \fig~\ref{fig:bias2} illustrates that the
presence of a star bursting galaxy enhances the probability and
efficiency of star formation, $f_*$, in nearby low-mass halos.
\fig~\ref{fig:bias2} is analogous to \fig~\ref{fig:bias1} but the blue
and magenta symbols show luminous halos with mass $<10^7$ M$_\odot$
with $f_*>10^{-3}$ and $f_*<10^{-3}$, respectively. Here,
$f_*=M_*/M_{bar}^{max}$, where
$M_{bar}^{max}=M_{dm}\Omega_b/\Omega_m$.  The yellow circles show
luminous galaxies with mass $>10^7$ M$_\odot$ and the black circles
show all dark halos with $M_{dm}>10^6$ M$_\odot$ as in
\fig~\ref{fig:bias1}.  The reason for dividing luminous galaxies in
two groups according to their mass will be evident in
\S~\ref{sec:stat_prop}. We will show that galaxies of the same mass but with
$M_{dm}<10^7$ M$_\odot$ can be either dark or luminous. Instead more
massive halos have a luminosity that is roughly increasing with the
halo mass (see \fig~\ref{fig:starf1}).  Galaxies with mass $<10^7$ M$_\odot$
are more likely to be luminous and have a high efficiency of star
formation (blue circles) if they are nearby other luminous galaxies
(\eg, yellow circles) than if they are isolated (\eg, magenta
circles). Even galaxies more massive than $10^7$ M$_\odot$ are more
likely to remain dark (\ie, large black circles) if they evolve
in isolation.

\section{Metal enrichment of the IGM}\label{sec:igm}

\def\capfigda{Volume filling factor of metal-enriched gas as a
  function of redshift for the simulations in \tab~\ref{tab:one}. Each
  panel, from top to bottom, shows the fraction of IGM volume with
  metallicity $\log \, (Z/Z_\odot) > -6, -5, -4$ and $-3$.
  (Left). These simulations are consistent with models of stellar
  reionization at $z \sim 6$. In these simulations metals are ejected
  due to photoevaporation from internal ionizing sources and tidal
  effects (run S1, S2 and S3) or, as in run S4, by tidal interactions
  only (long-dashed curves).  Simulation S1 (solid curves) is our
  high-resolution run, S2 and S3 are lower-resolution runs with weak
  and strong radiative feedback, respectively. Mechanical energy input
  from SN explosions is not included. (Right) These simulations have
  typically earlier epoch of reionization, consistent with WMAP-1 and
  WMAP-3.  Here, metals are ejected only due to tidal effects, as in
  run S6 (dotted curves), by photoevaporation due to X-ray heating, as
  in run S5 (dashed curves), or by strong SN feedback, as in run S7
  (solid curves).}  \placefig{
 \begin{figure*}[tbhp]
 \epsscale{1.1}
\plottwo{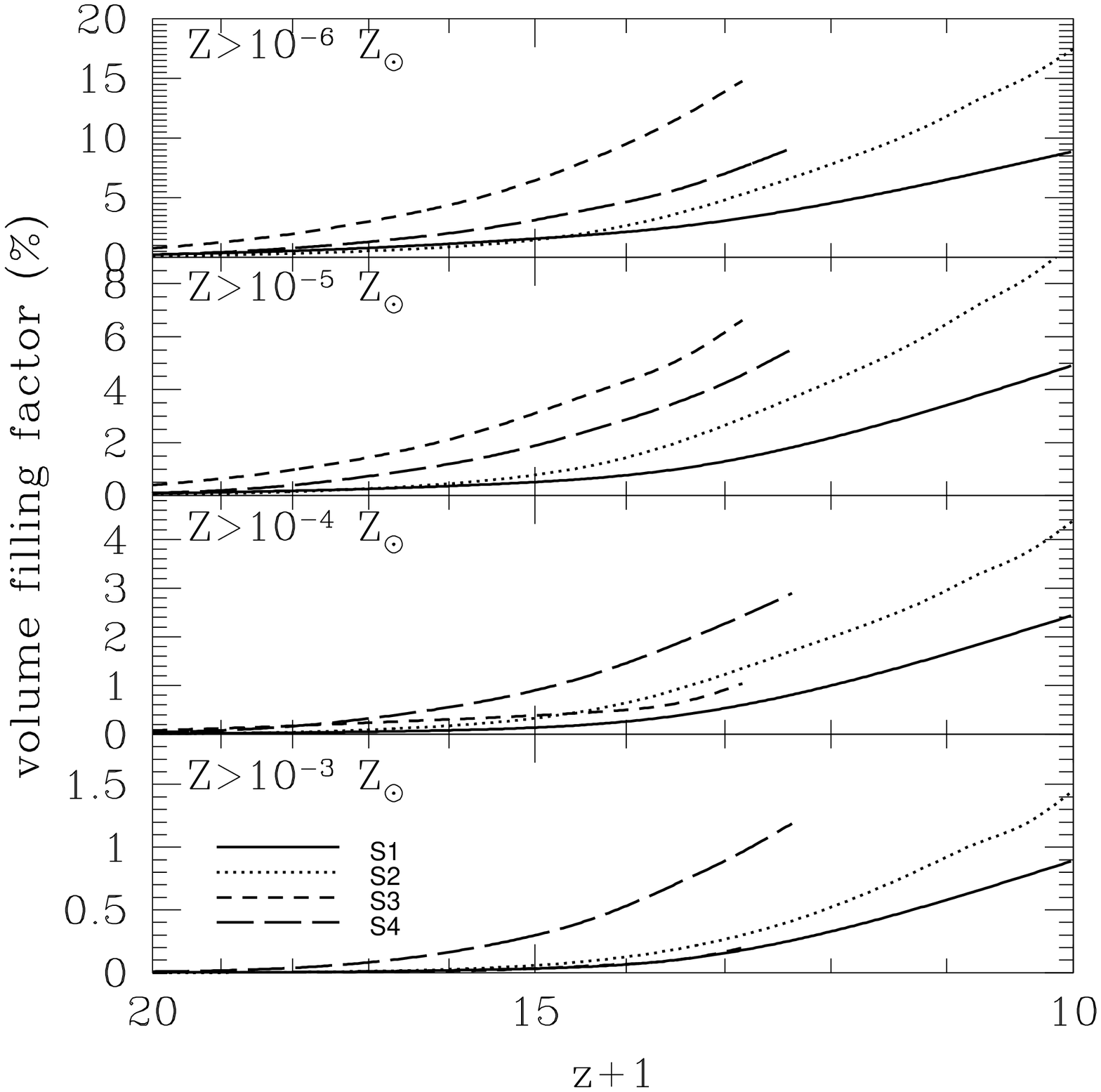}{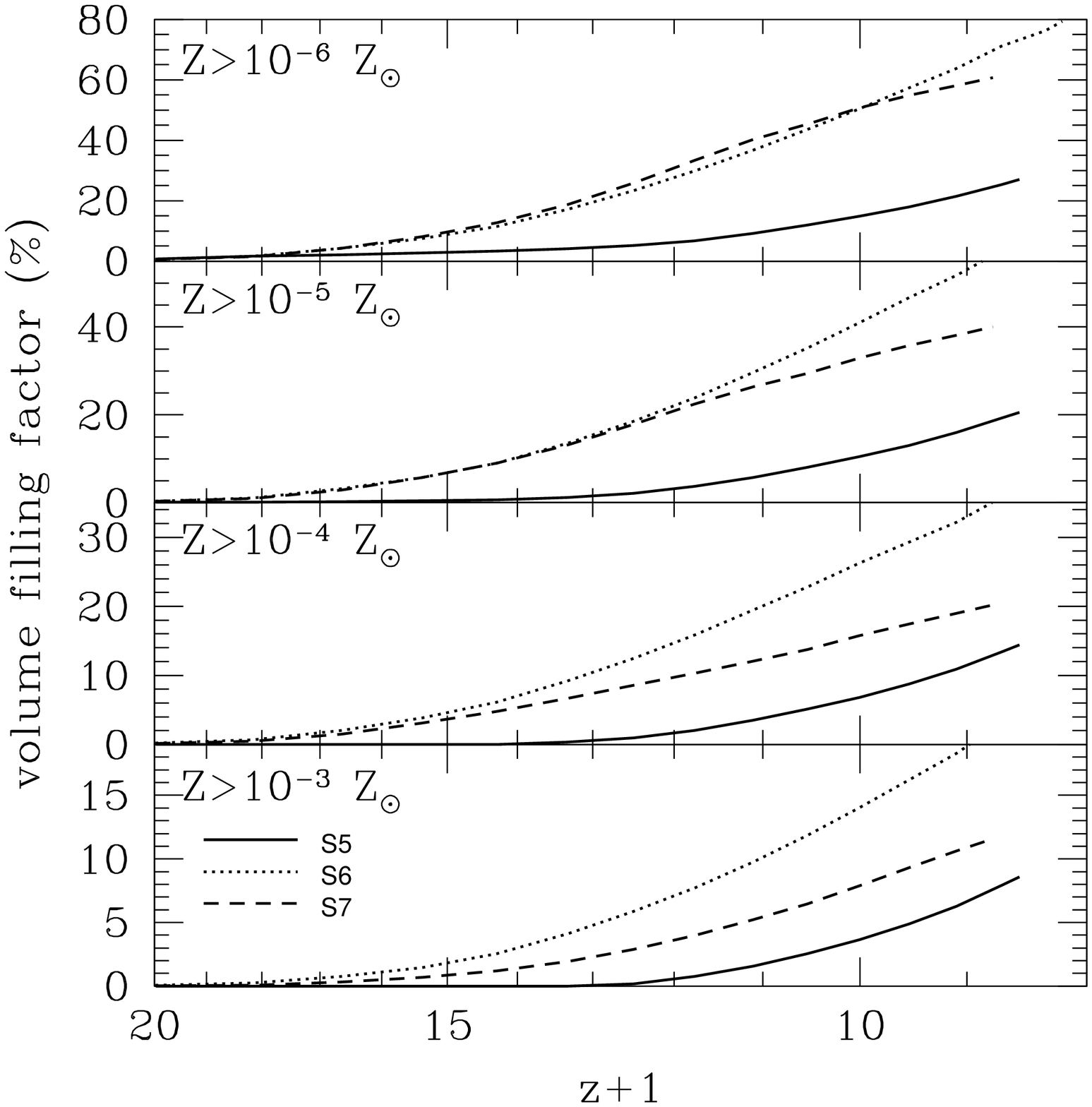}
 \caption{\label{fig:fill2}\capfigda}
 \end{figure*}
}

Metallicities of about $Z \sim 10^{-3}$ Z$_\odot$ are measured from
observation of absorption lines in the Ly$\alpha$ forest at $z \sim 2-5$
\citep[\eg,][]{Songaila:01, Schaye:03, Pettini:03, Simcoe:04}. This
mean metallicity, typically inferred from abundances of \ion{C}{iv}
and \ion{Si}{iv}, shows little evolution from $z=5$ to $z=2$.  However,
it is not well understood if a possible redshift-dependent ionization
correction may conspire in hiding a real metallicity evolution.  The
homogeneity of the metal distribution in the IGM is also unknown. This
is an important measure, because it can be used to determine the
properties of the galaxies responsible for the observed metal
enrichment.  If underdense regions of IGM contain some metals, then
star formation in high-redshift galaxies, of the type studied in this
work, may have pre-enriched it \citep[\eg,][]{MadauF:01}. If,
instead, the metals detected along a line of sight are associated with
nearby bright galaxies at $z \sim 2-5$, the metal distribution is
probably inhomogeneous. In this second case, observations are
not probing a minimum floor of metal enrichment produced by the first
galaxies.  The current observational sensitivity to metallicity in
high-redshift absorbers is
$\sim10^{-3.5} Z_{\odot}$ \citep{Songaila:01, Simcoe:04}.

In this section, we analyze the contribution to the IGM metallicity by
low-mass galaxies at $z > 8$.  We show that the metals produced by the
first galaxies can fill the space between bright galaxies uniformly
only if we make extreme assumptions on the strength of SN feedback
(\ie, in run S6 and S7) and only to very low metallicities floors
($Z/Z_\odot \simlt 10^{-5}$). For less extreme cases, the volume
filling factor is typically $<10$\% even at small metallicity floors.
Thus, the first galaxies may not able to pollute with metals all the
IGM volume, living some large underdense regions of 0.5 comoving Mpc
in size (at $z \sim 10$), with primordial composition.

The volume filling factor of the IGM enriched to the typical
metallicities observed in the Ly$\alpha$ forest is less than 1\% in
most runs. It is therefore unlikely that the metal absorption systems
associated with the Ly$\alpha$ forest were produced by the very first
low-mass galaxies.

In our simulations three different processes can be important in
polluting the IGM with metals: (i) metals ejected by galaxy
harassment, as proposed by \cite{Gnedin:98a}; (ii) by galactic winds
produced by photoionization in low-mass galaxies (Paper~2); and (iii)
by SN explosions \citep[\eg,][]{Ferrara:00, MadauF:01,
Fujita:04}. Using a set of simulations, we can analyze separately the
contribution from each process. Before showing the results, we should
mention a few caveats. The metal ejection from SN explosions is model
dependent, since it is not possible to resolve the multi-phase
structure of the ISM and SN shock fronts.  We model SN feedback as in
\cite{Gnedin:98a}: energy and metals injection from SNe is local
(within the resolution element). We assume perfect mixing of the
metals within each cell (\ie, the sub-grid multi-phase medium has
homogeneous metallicity). The energy injection from SNe contributes to
the heating of the sub-grid multi-phase medium and to increase its
velocity dispersion. These terms are calculated using analytical
solutions of shocks propagating in a medium with temperature, pressure
and density equal to the mean values for the cell. The pressure of the
gas includes two terms, for thermal and turbulent pressure.  Hence, of
the three processes that eject metals into the IGM, only (i) and (ii) 
are independent of ``sub-grid'' physics.
Finally, the total mass of metals produced and the volume filling
factor of gas with metallicity larger than a floor, $Z=Z_0$, is
directly proportional to the assumed yields of the stellar population.

In \fig~\ref{fig:fill2} we show the volume filling factor of
metal-enriched gas as a function of redshift for a set of simulations
listed in \tab~\ref{tab:one}. Each panel, from top to the bottom,
shows the volume filling factor of the IGM, $f_V(Z>Z_0)$, that is
enriched to metallicities larger than a ``floor", $Z_0$, taken
to be $5\times 10^{-6}$, $5\times 10^{-5}$, $5\times 10^{-4}$, and
$5\times 10^{-3}~Z_\odot$. In all our simulations, about $10\%$ of all
metals produced end up in the IGM, and the rest resides in galaxies
and stars. The metallicity is larger than the mean in overdense
regions and lower than the mean in underdense regions. We should be
cautious in comparing the results of the high-resolution simulation
with the lower resolution ones, as the metal filling factor may be
affected by the numerical resolution.  In addition, because of the
small box size of the simulations, these metal filling factors become
increasingly inaccurate at low redshifts due to the missing clustering
on large scales. But galaxy clustering that we miss in will only make
the filling factors smaller. Hence the values we calculate should be
treated as upper limits.

In the left panels, we show a set of simulations from Paper~2,
consistent with models of stellar reionization at $z \sim 6$. In all
these simulations, we neglect the mechanical feedback from SN
explosions; therefore the metal ejection from the galaxies is produced
by tidal stripping of gas or photoevaporation from internal sources,
or both.  The long-dashed lines refer to run S4 without radiative
feedback.  All the other lines show runs that include radiative
feedback; run S2 with $\langle f_{esc}\rangle=1\%$ (weak radiative
feedback - dotted lines) run S3 with $\langle f_{esc}\rangle=100\%$
(strong radiative feedback - short-dashed lines), and high-resolution
run S1 with $\langle f_{esc}\rangle=10\%$ (solid line).

In the right panels, we show a set of simulations from RO04 and ROG05,
consistent with early IGM reionization suggested by WMAP-1 (optical
depth to Thompson scattering $\tau_e \approx 0.17\pm0.04$).  However,
WMAP-3 data \citep{Spergel:07} imply a lower optical depth, $\tau_e =
0.09 \pm 0.03$, which may be explainable without large amounts of
high-$z$ star formation \citep{ShullVenkatesan:07}.  Some of these
simulations also include strong mechanical feedback from SN
explosions. In particular, the dotted lines show a simulation with an
early X-ray partial ionization and reheating. In this simulation
metals are dispersed in the IGM mainly from the photoevaporation of
low-mass galaxies. The solid and dashed lines show simulations with
early reionization from \pop3\ stars with top-heavy IMF and $\langle
f_{esc}\rangle=0.5$ (\ie, strong radiative feedback). The solid lines
show a case where we also include strong feedback from SN explosions
(\eg, \pop3\ stars, having typical masses in the range $100-300$
M$_\odot$, some of which end their lives as pair-instability SNe).
The dashed line shows a case in which energy input from SN explosions
is negligible, such as \pop3\ stars with masses $<140$ M$_\odot$ or
$>260$ M$_\odot$, that end their lives collapsing into BHs without
energetic SN explosions or without exploding at all.

The results show that low-mass galaxies are more effective than
massive ones in enriching a large volume fraction of the IGM, but only
to very low-metallicities.  Supernova explosions, tidal stripping of
metals, and photoevaporation of low-mass galaxies have similar
importance for transporting metals from galaxies to the low density
IGM.

We interpret the results shown in \fig~\ref{fig:fill2} with
the aid of a toy model, fitting the model to the simulation results. In
order to calculate the volume of metal enriched gas, we integrate the
contribution of each galaxy as a function of time to find the IGM porosity,
\begin{eqnarray}
Q_V(Z>Z_0 ,t)&=&{4\pi \over 3}\int_0^t dt^\prime \int_0^\infty {d
M_{\rm dm} \over M_{\rm dm}}\\
&\times& {\partial \over \partial t^\prime} \, [n_{\rm gal}(M_{\rm dm},t^\prime)
R_{\rm met}^3(Z_0,M_{\rm dm},t^\prime)] \; ,\nonumber
\label{eq:qv}
\end{eqnarray}
where $0<Q_V<\infty$ is the porosity, $n_{\rm gal}(M_{\rm dm},t)$ is
the volume number density of galaxies of total mass $M_{\rm dm}$, and
$R_{\rm met}(Z_0,M_{\rm dm},t)$ is the radius of metal-enriched gas
with metallicity $Z>Z_0$ around each galaxy.  The volume filling
factor is related to the porosity by the relationship $f_V \equiv
1-\exp[-Q_V]$, so that $f_V \simeq Q$ if $Q \ll 1$, and $f_V=1$ as
$Q_V \rightarrow \infty$. Due to the effect of galaxy clustering this
simple model may break down for relatively small values of $f_V$. The
interpretation with our toy model of the results shown in the upper
panels of \fig~\ref{fig:fill2}, for which $f_V>1\%$, may be
significantly inaccurate.
  
For fixed $Z_0$, we find that the porosity increases with time as
$Q_V(t) \propto t^3$ within 20\% error.  Within the same error, the
porosity of metal-enriched IGM with $Z>Z_0$ as a function of time is
well approximated by the fitting formula,
\begin{equation}
   Q_V(Z>Z_0) \simeq A \left({Z_0 \over 0.05 Z_\odot}\right)^{-\alpha} 
    \left[ t \over t_{z=9} \right]^3 \; ,  
\end{equation}
where $t_{z=9}$ is the Hubble time at $z=9$ and the parameters $A$
and $\alpha$ (\tab~\ref{tab:two}) depend on the metal yield, the IMF,
and the feedback processes included in the simulations.  

The mean physical distance between small mass galaxies at high
redshift is only a few kpc. In principle, these small galaxies may
pollute the IGM quite uniformly with metals. However, the smaller the
mass of the galaxy, the lower its ability to form stars and produce
metals. For this reason, the metals ejected into the IGM by the more
numerous population of small mass galaxies have a larger feeling
factor than metals produced by more rare massive galaxies, but can
only enrich the IGM to a very low metallicity floor.

\def\tabtwo{
\begin{deluxetable}{llcccc}
\footnotesize
\tablecaption{Parameters: Volume Filling Factor of Metal-enriched Gas.
\label{tab:two}}
\tablewidth{0pt} 
\tablehead{ \colhead{Name} & \colhead{RUN}
    &\colhead{$A$} & \colhead{$\alpha$} & \colhead{$\langle Z/Z_\odot \rangle_M$} &
    \colhead{$\langle Z/Z_\odot \rangle_V$}} 
\startdata
  S1 & 256L1p3     & 1.0 & 0.35 & -2.84 & -4.32 \\
  S2 & 128L1p2-2   & 1.9 & 0.35 & -2.95 & -4.19 \\
  S3 & 128L1p2f1   & 1.6 & 0.50 & \nodata & \nodata \\
  S4 & 128L1noRAD  & 3.4 & 0.30 & -2.31 & -3.56 \\
  S5 & 128L1XR & 3.0 & 0.25 & -3.71 & -5.145 \\
  S6 & 128L2BH & 5.0 & 0.35 & -2.82 & -4.056 \\
  S7 & 128L2PI & 8.0 & 0.30 & -2.28 & -3.542 
\enddata
\end{deluxetable}
}
\placefig{\tabtwo}

We could assume, for simplicity, that galaxies of mass $M_{\rm dm}>M_0$ 
are responsible for the enrichment to a metallicity $Z > Z_0$.  
For galaxies of comoving spatial density $n_{\rm gal}$, the mean 
physical distance between star-forming galaxies is 
$d_{\rm gal} \approx n_{\rm gal}^{-1/3}/(1+z) \sim$ 1--30~ kpc, depending 
on their mass (see \S~\ref{sec:obs}). Given that the volume filling factor 
is between a few percent to 50\% depending on $Z_0$, we estimate from 
\eq~(\ref{eq:qv}) that $R_{\rm met} \sim 1$ kpc. Metal-enriched gas expanding
at constant speed $v_{\rm ej}$ for a time $t$ travels a distance,
\begin{equation}
R_{\rm met}= (1 ~{\rm kpc}) \left({v_{\rm ej} \over {\rm 10~km~s}^{-1}}\right) \left({t \over {\rm 100~Myr}}\right).
\label{eq:ej}
\end{equation}
Since the Hubble time at high redshifts ($10<z<30$) is $t_H \sim
0.1-0.5$ Gyr and the bursting star formation has time scales 
$t_{\rm burst} \sim 10$ Myr, this distance requires that 
$10 < v_{\rm ej} < 100$ km s$^{-1}$. 
The lower bound of $v_{\rm ej}$ is about the escape velocity from
dPri galaxies and is consistent with metal pollution by processes (i)
and (ii). Larger values of $v_{\rm ej}$ could be produced by SN
explosions, but from \eq~(\ref{eq:ej}) we infer that these higher
velocities are in place for only a fraction of the Hubble time at $z
\sim 10$.

We note that, if our simulation volume at $z \sim 8-9$ becomes a mean
density or an underdense region of the universe at $z \sim 2-5$, the
clustering and galaxy formation will be slow from $z\sim 8-9$ to $z
\sim 2-5$, and the filling factor of metal-enriched gas should not
increase considerably.  The metallicity in overdense regions at $z
\sim 2-5$ will be produced mostly by newly formed galaxies not
included in our simulations.

\section{Population of Dwarf Primordial Galaxies}\label{sec:stat_prop}

In this section we analyze the properties of the simulated population of 
dPri galaxies (their stellar and gas content as a function of total mass)
and the internal properties of individual objects (gas and stellar density 
profiles).

\subsection{Statistical Properties of the Galaxies}

\def\capfigdb{Fraction of baryons (sum of gas and stars) to dark
  matter (normalized to the cosmic mean value $f_{\rm bar}^{\rm max}=
  \Omega_b/\Omega_m=0.136$) as function of the halo mass $M_{\rm dm}$
  for run S1 at $z=10$ . The size of the dots is proportional to the
  fraction of stars $f_*=M_*/M_{\rm bar}^{\rm max}$ in each halo: from
  the largest to the smallest dots we have $f_*>10$\%, $1<f_*<10$\%,
  $0.1<f_*<1$\% and $f_*<0.1$\%, respectively. The baryon content in
  dark halos (smallest dots) smaller than a few $10^6$ M$_\odot$ drops
  sharply because of the finite IGM temperature (the mass cut of
  indeed is at smaller masses at earlier times when the IGM
  temperature is lower). In low-mass halos that form some stars, the
  baryon fraction is further decreased by photoheating from internal
  sources of radiation.}  
\placefig{
 \begin{figure}[tb]
 \epsscale{1.1}
\plotone{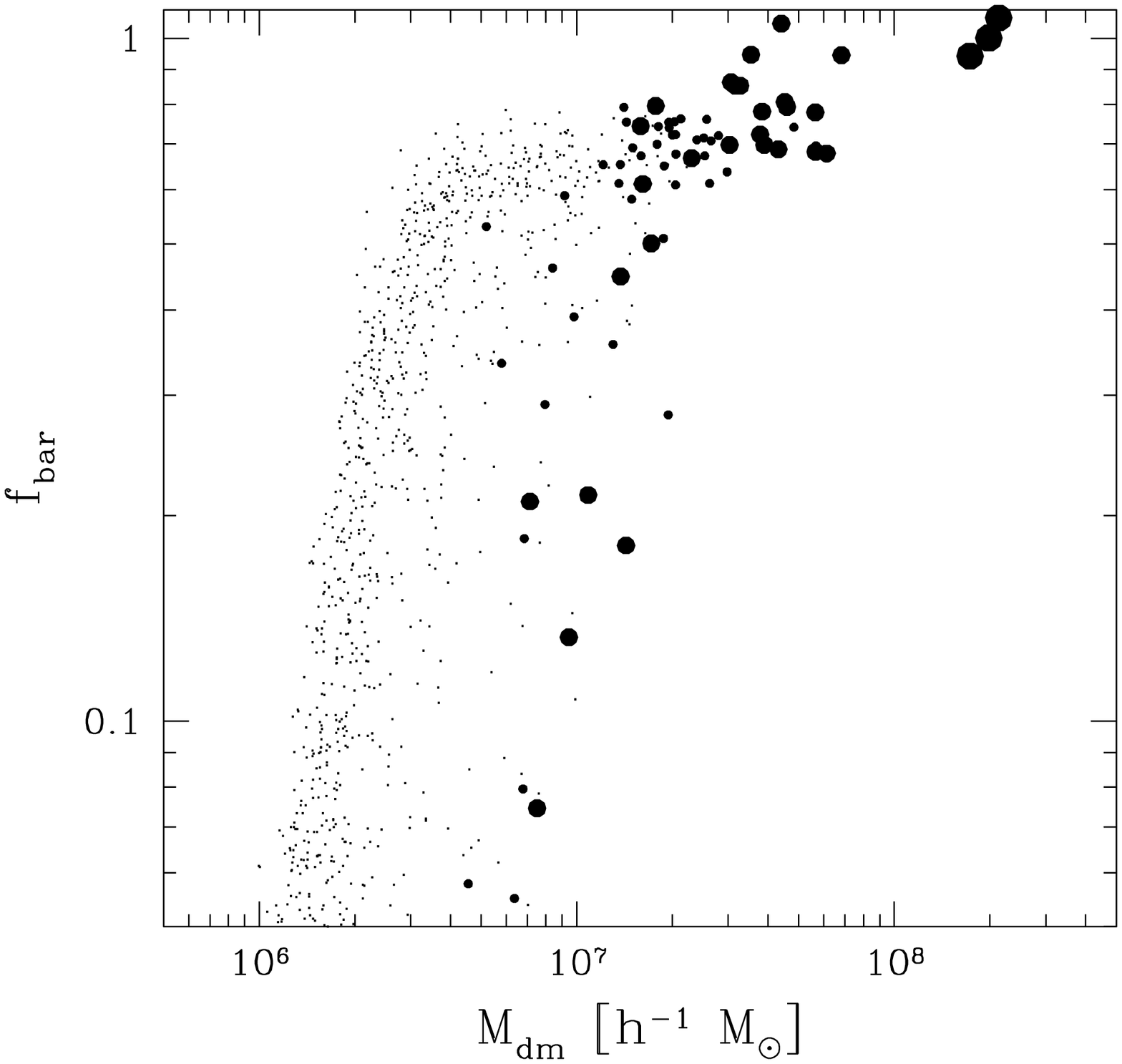}
 \caption{\label{fig:bar1}\capfigdb}
 \end{figure}
} 
\def\capfigea{Fraction of gas converted into stars as function of
halo mass of the galaxy for S1 run (left) and S2 (right) at $z=10$.
Circles, from smaller to the larger, refer to galaxies with gas
fractions $f_g<0.1$\% (blue), $0.1\%<f_g<1$\% (cyan), $1\%<f_g<10$\%
(red) and $f_g>10$\% (green), respectively. There is a clear lower
envelope for the star formation efficiency, $f_*$, in halos with
masses $>10^7$ M$_\odot$, roughly proportional to the halo mass. The
scatter of the values of $f_*$ increases with decreasing halo mass
and, remarkably, the upper envelope of $f_*(M_{\rm dm})$ is almost
independent of the halo mass. This reflects the fact that local
feedback plays an important role in determining the luminosity of
low-mass galaxies.}  
\placefig{
\begin{figure*}[tb]
\epsscale{1.1}
\plottwo{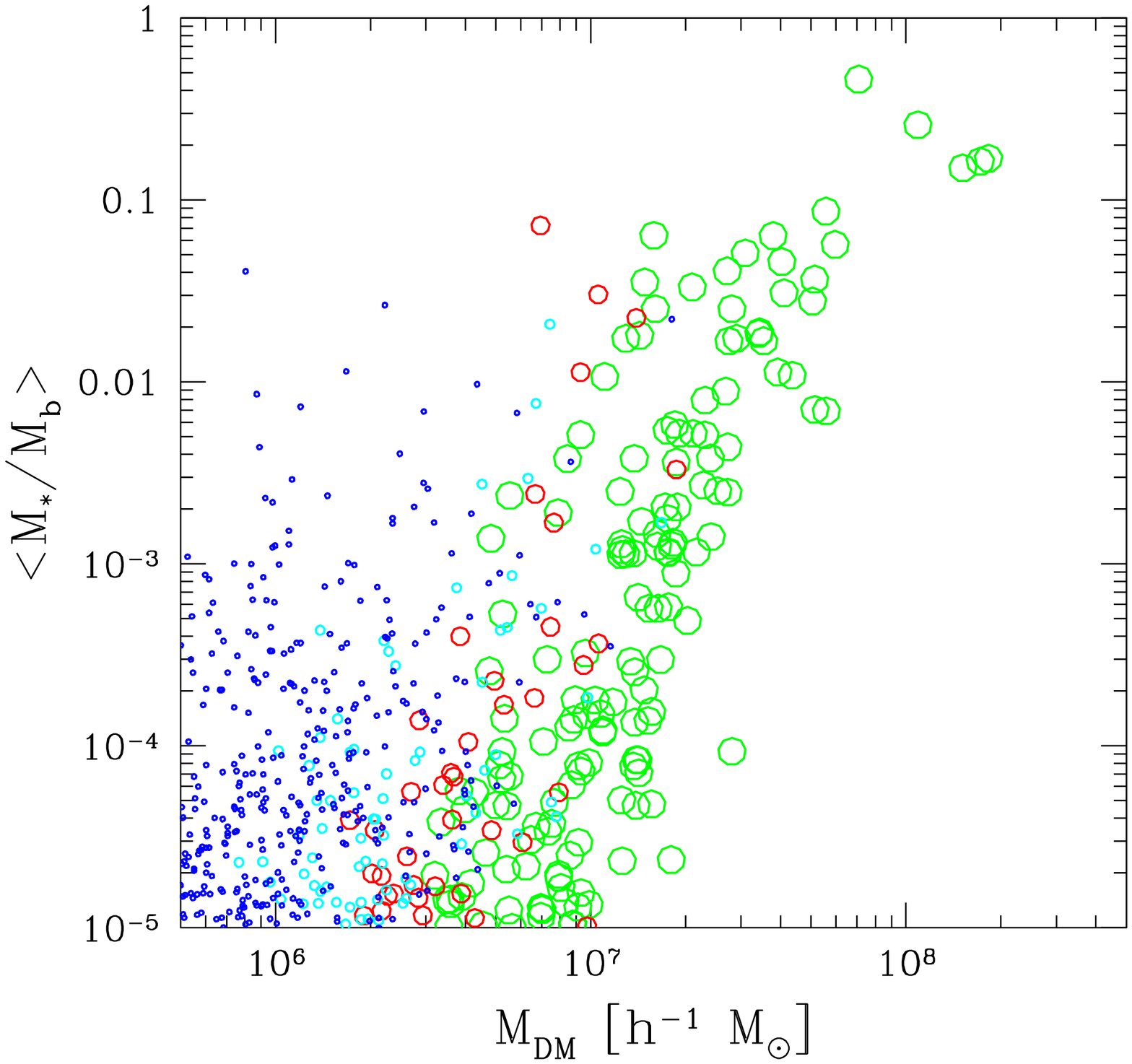}{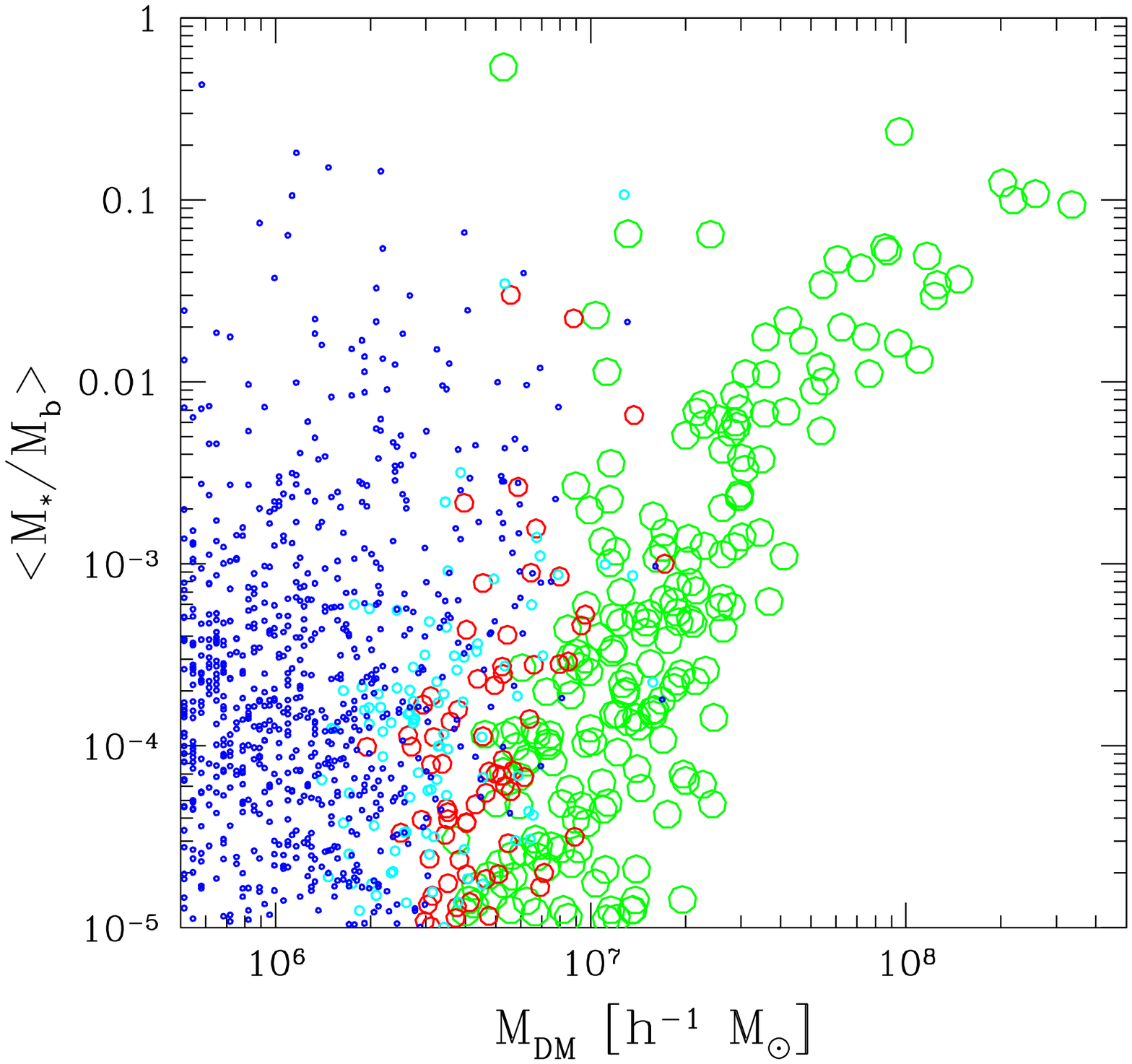}
\caption{\label{fig:starf1}\capfigea}
\end{figure*}
} 
One of the most distinctive properties of simulated dPri galaxies is
their low baryon-to-dark matter ratios. This is not too surprising,
since in most of our simulations we include mechanical energy input
from SN explosions.  Most of the intergalactic gas that falls into the
deep gravitational potential of massive galaxies and clusters is
effectively trapped.  In this case, it is a good approximation to
assume that their total baryonic mass (the sum of gas and stars) is
$M_{\rm bar}^{\rm max} \simeq (\Omega_b/\Omega_m) M_{\rm dm}$, where
$\Omega_b/\Omega_m\simeq 1/6$ is the cosmic mean of the baryon-to-dark
matter mass ratio.  In \fig~\ref{fig:bar1} we show the fraction of
baryons in each simulated galaxy normalized to the cosmic mean,
$f_b=(M_*+M_g)/M_{\rm bar}^{\rm max}$, as a function of their total
mass $M_{\rm dm}$. The figure refers to run S1 at $z=10$. The size of
each circle represents the fraction of stars $f_*=M_*/M_{\rm bar}^{\rm
max}$ in each galaxy.  As expected, more massive galaxies retain on
average a larger fraction of their initial baryon content, but the
scatter of the $f_{\rm bar}-M_{\rm dm}$ relationship increases with
decreasing halo mass.  The baryon content in star-free halos (smallest
dots) depends on the IGM temperature. When we examine the same plots
as in \fig~\ref{fig:bar1} but at redshifts $z \simgt 13$ we find $f_*
\sim 1$. As the temperature of the IGM increases, the baryon fraction
diminishes, first in the smaller halos and then in the larger ones.
In those halos that form some stars, the baryon fraction is further
reduced by photoevaporative winds produced by the internal sources of
radiation. As a result, the distribution of the baryon fraction inside
halos as a function of their mass depends on whether they are dark or
luminous: the typical cut off mass decreases with decreasing $f_*$ and
eventually equals the minimum cut off mass set by the IGM temperature.
\def\capfigeb{{\it
(Left)}. Average stellar fraction $\langle f_* \rangle$ (thick curves),
and gas fraction $\langle f_g \rangle$ (thin curves), as a function of
halo mass at $z=14.6, 12.5$, and $10.2$ for the run S1.  {\it
(Right)}. Same as the left panel, for sun S2 at $z=12.5, 9.6$, and
$9$. For comparison, the symbol with error bar shows the expected star
formation efficiency in the first mini halo of mass $10^6$ M$_\odot$
simulated by \cite{Abel:02}.}  \placefig{
\begin{figure*}[tb]
\epsscale{1.1}
\plottwo{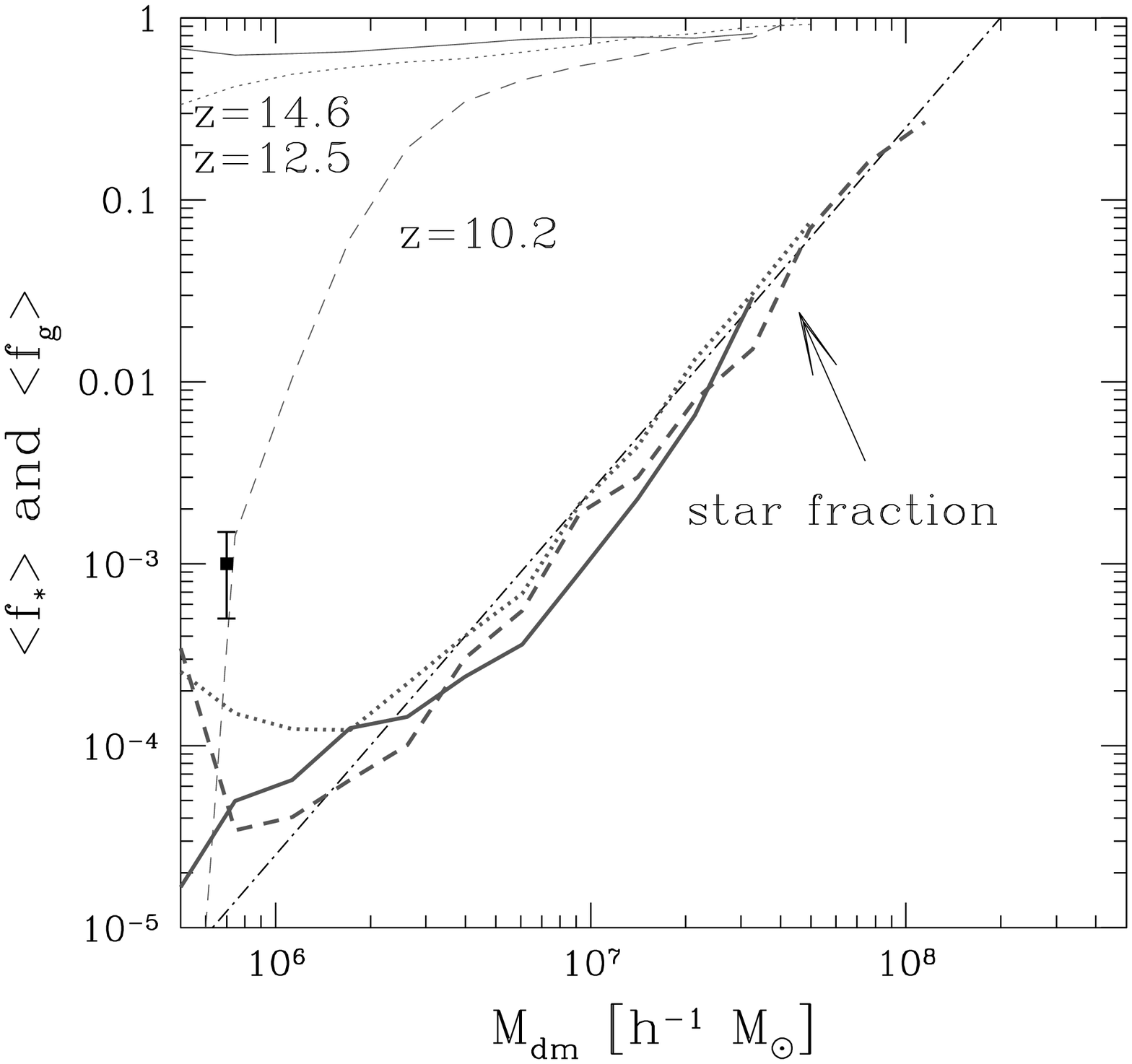}{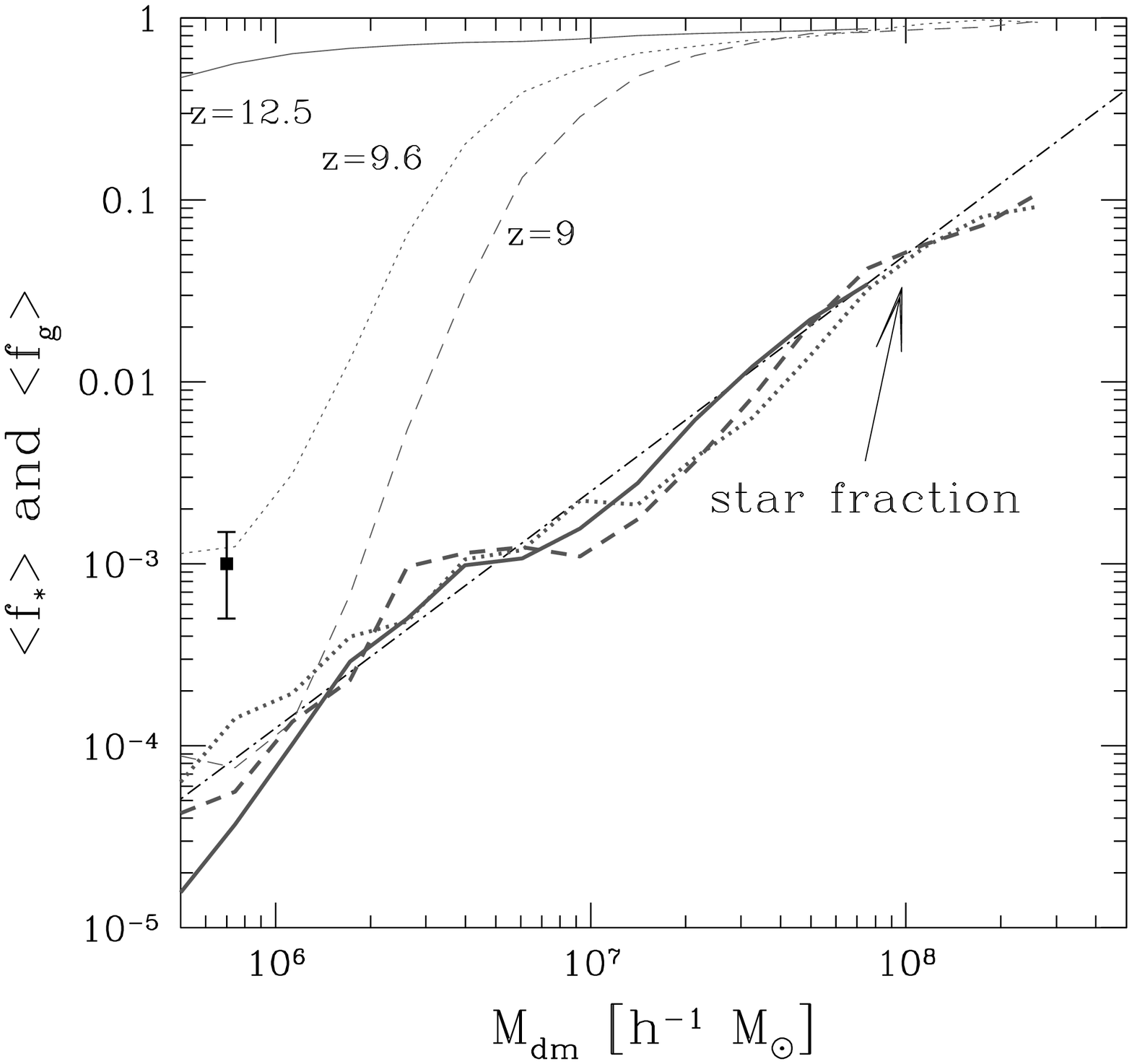}
\caption{\label{fig:starf2}\capfigeb}
\end{figure*}}

The large variations of the mass-to-light ratio and the gas fraction
in halos of identical total mass is an indication of the local nature
of the feedback processes.  This is illustrated in
\fig~\ref{fig:starf1} that shows the stellar fraction $f_*=M_*/M_{\rm
bar}^{\rm max}$ as a function of the halo mass, $M_{\rm dm}$, at
$z=10$ for the run S1 (left figure) and run S2 (right figure). Each
galaxy is shown with symbols of different sizes proportional to their
gas fraction. Halos of masses $M_{\rm dm}<10^7$ M$_\odot$ can be
completely dark (most of them), while low luminosity galaxies and a
few brighter galaxies can have 10--50\% of their gas converted into
stars. The scatter is much reduced for galaxies with masses $M_{\rm
dm}>10^7$ M$_\odot$ that retain most of their gas, and their stellar
fraction shows a much tighter relation to their total mass.  At $z
\sim 10$, galaxies with $M_{\rm dm}>10^7$ M$_\odot$ are still gas
rich, with a subdominant stellar component.  The main physical
processes responsible for the low efficiency of star formation are
photoevaporation from internal sources and global feedback (\eg,
photodissociation of H$_2$ and IGM reheating).  

The mean values of the stellar and gas fractions as a function of the
halo mass are shown in \fig~\ref{fig:starf2}. In the left panel,
we show the mean as a function of DM mass at $z=12.5, 9.6$, and
$9$ for run S1, and the right panel shows the same for run S2.  
Dwarf galaxies that form stars show large variations in their gas
content because of stellar feedback and photoionization. The
gas is photoevaporated first from low-mass galaxies and as time
progress in larger ones.  Luminous galaxies with $M_{\rm dm}<10^8$ 
M$_\odot$ lose most their gas well before reionization and
star formation is halted.  New star formation cannot take place in
those small halos unless their mass increases as a result of
subsequent minor mergers.

We find that the mean star-formation efficiency 
$\langle f_* (t) \rangle = \langle M_*/M_{\rm bar, max} \rangle$ in a halo 
of mass $M_{\rm dm}$, is nearly time-independent and is well approximated 
by a power law 
\begin{equation}
\langle f_* \rangle(t) \simeq \epsilon_* \left({M_{\rm dm} \over 
    10^8~{\rm M}_\odot}\right)^\alpha,
\label{eq:mtl}
\end{equation}
where $\epsilon_*$ is the assumed star-formation efficiency. There is
a weak dependence of the exponent in the power law on the strength of
the feedback: $\alpha=1.5$ if the feedback is weak (run S2) and
$\alpha=2$ if the feedback is stronger (run S1).  
However, contrary to what happens to the gas fraction, the stellar
fraction does not evolve significantly with time.
Equation~(\ref{eq:mtl}) can be an useful approximation in
semi-analytic models for galaxy formation.  The point shown with a
cross in \fig~\ref{fig:starf2} is the first galaxy simulated with very
high resolution by \cite{Abel:02}. The point lies above the mean value
for galaxies with mass $M=10^7$ M$_\odot$. This is not surprising,
because in our simulations most of the galaxies are dark, owing to
feedback processes not included in simulations of the first stars.
\def\capfigfa{({\em Left}). Mass fraction of stars as a function of
  their metallicity $Z$ in solar units. Each panel refers to a
  different simulation, where solid, dotted and dashed lines show 
  mass at $z=17.5, 14.6$ and $12.5$, respectively. In (a) we show 
  run S4, in (b) run S2, in (c) run S3, and in (d) run S1. From top 
  to bottom, feedback is stronger and the global mode of star formation 
  changes from continuous to bursting. ({\em Right}). Same as left figure, 
  but showing cumulative mass in stars with metallicity lower than $Z$.}  
\placefig{
\begin{figure*}[tbhp]
\epsscale{1.1} 
\plottwo{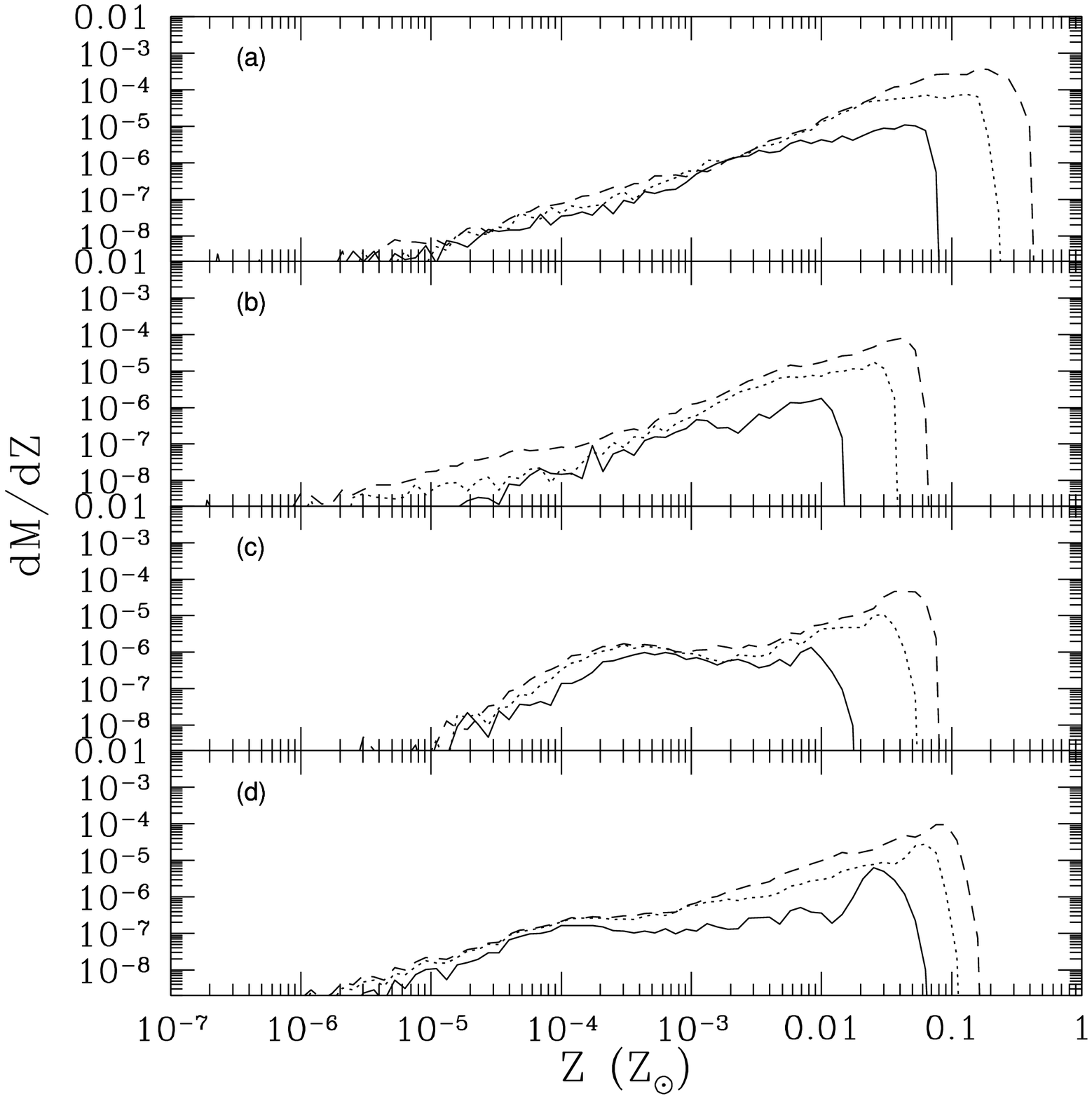}{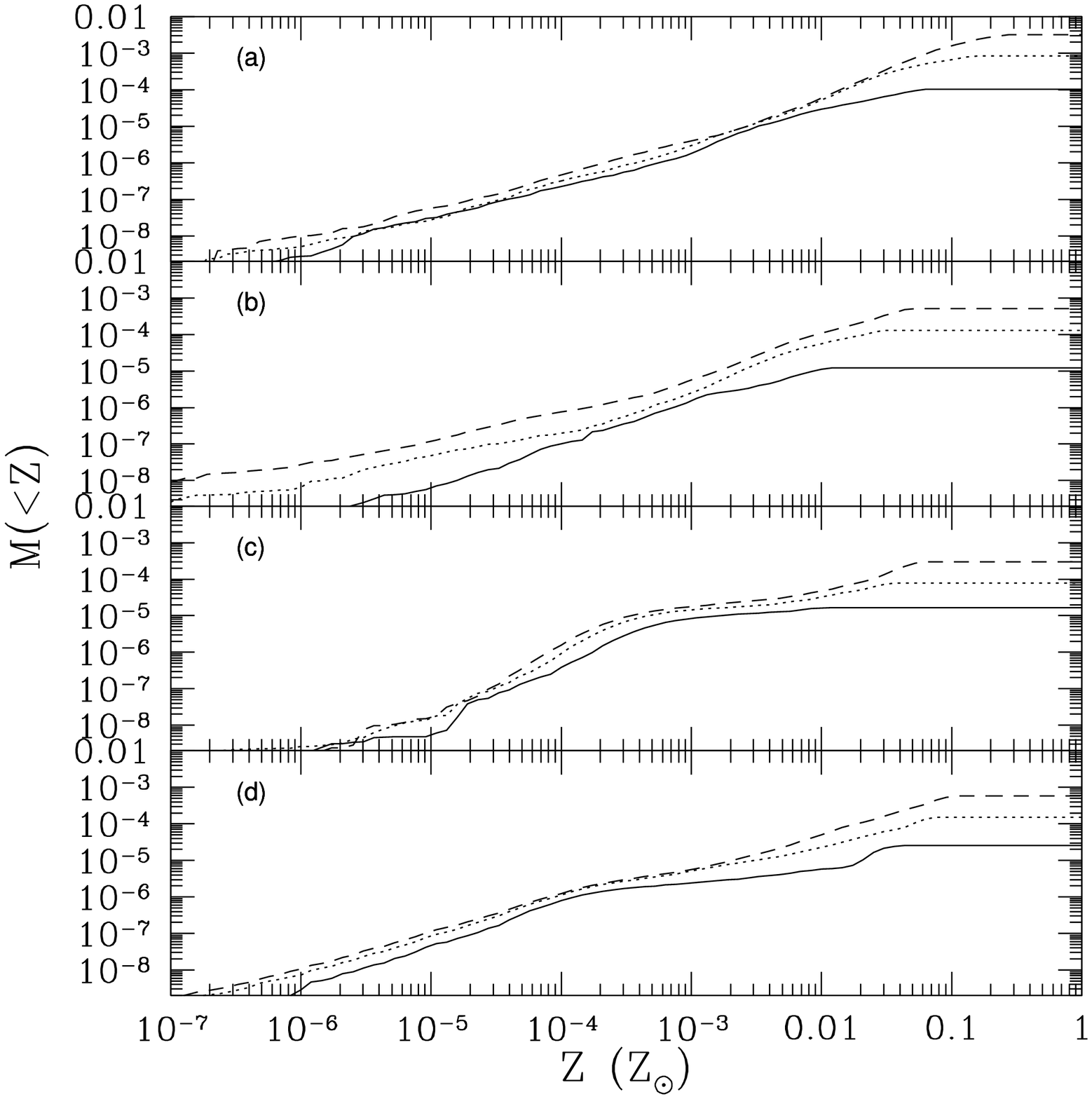}
\caption{\label{fig:zmasf1}\capfigfa}
\end{figure*}}

\subsection{Metallicity Distribution of the Stars}\label{ssec:smet}

In this section we study the metallicity distribution of the first
stars in the simulations from Paper~2.  First, we clarify
the numerical limitations of the simulations, which do not attempt to 
model the mechanical energy input from SNe. The resolution of the simulation 
is insufficient to resolve the complexity of physical processes that
regulate the physics of the multi-phase ISM.  In each resolution element,
we assume perfect metal mixing, although it is possible that the metal
distribution is inhomogeneous on smaller scales.  Thus, pockets of
zero-metallicity gas may survive even if the gas has a large mean
metallicity.  As discussed in \S~\ref{sec:igm}, in those simulations
where we have also included SN feedback, we find, in agreement with
\cite{Gnedin:98a}, that mechanical feedback does not affect star
formation in galaxies with $M_{\rm dm} \simgt 10^8$ M$_\odot$. In lower
mass galaxies, when we include the effect of SNe, the results are
similar to simulations with strong photoevaporative galactic winds
discussed in this section.  Therefore, given the uncertainties
introduced by the model-dependent treatment of SN explosions, we
prefer to study only the cases that include the better understood
radiative feedback.

Given the aforementioned caveat, \fig~\ref{fig:zmasf1}~(left) shows
the mass fraction of stars as a function of their metallicity $Z$ in
solar units. Each panel refers to a different simulation, at $z=17.5,
14.6$, and $12.5$.  In \fig~\ref{fig:zmasf1}~(right) we show the
cumulative mass in stars with metallicity lower than $Z$ expressed in
solar units.  In each panel, from top to bottom, we show the
metallicity distribution in simulations with the following properties:

\noindent
{\it Panel (a): no feedback} - star formation is continuous and there
are no galactic winds due to SN explosions or photoionization.  Some
enriched gas might be lost from interacting galaxies by tidal
stripping during galaxy interactions or mergers \citep{Gnedin:98a}.
Metals can be accreted by larger galaxies when smaller satellites fall
into them. In this simulation, the low-metallicity stars with $Z<0.01$
Z$_\odot$ stop forming at redshift $z \sim 17$, and the mean metallicity
of stars increases with time. By $z=10$ most stars have 
$0.05 < Z/Z_\odot< 0.5$.

\noindent
{\it Panel (b): weak feedback} - star formation is continuous and
takes place also in very low-mass galaxies but with low efficiency.  
The formation of stars in these smaller galaxies is delayed, and 
zero-metallicity stars may continue to form at $z \sim 10$.

\noindent
{\it Panel (c): strong feedback} - bursty star formation takes
place only in higher mass galaxies and is suppressed in the smaller 
ones.  There are fewer extremely low metallicity stars
($Z/Z_{\odot} < 10^{-5}$) and an uniform distribution between 
$Z/Z_{\odot} \sim 10^{-4}$ to $10^{-2}$.  After redshift $z\sim10$, 
as low-metallicity stars stop forming, star formation takes 
place only in pre-enriched gas.

\noindent
{\it Panel (d): intermediate feedback and higher resolution} - in this
simulation the metallicity distribution of stars is an intermediate case
between panel (b) and (c).

In order to estimate the number of low metallicity stars expected in
the present day Universe from the data in our simulations, several
assumptions need to be made. Here, we briefly describe how
one should proceed in order to estimate the number of low-metallicity
stars in our Galactic halo. Given our rather simplistic treatment,
these numbers should be considered order of magnitude
estimates. More sophisticated modeling could be adopted to compare
the simulations to detailed observational data \citep{Tumlinson:06}.

First, we note that even at redshifts $z \sim 10-20$ most stars have
metallicities of about 1/10 solar, while only $1$\% have ultra low
metallicities ($Z \simlt 10^{-3}~Z_{\odot}$). Let's assume that ultra
low metallicity stars stop forming sometime before redshift $z \sim
10$: as it seems to be the case in all our simulations but S2, the one
with weak radiative feedback. Using a rough estimate of the mass of
stars that forms in the Universe after $z=10$, we estimate a fraction
of stars (in mass) in the present universe with metallicity $Z <
10^{-4}~Z_{\odot}$ of roughly $10^{-6}$.\\ The stars in our Galactic
halo have total mass of about $10^{10}$ M$_\odot$.  Therefore,
assuming a uniform mixture of low and higher metallicity stars, a mass
of about $10^4$ M$_\odot$ of our Galactic halo should be in stars with
$Z < 10^{-4}~Z_{\odot}$. Roughly, in our simulations, the number of
low-metallicity stars decreases by a factor of ten, going down a
decade in metallicity (see \fig~\ref{fig:zmasf1}).  Thus, for stars
with $Z<Z_{cr}$ where $Z_{cr}=10^{-6}$, $10^{-5}$, $10^{-4}$, and
$10^{-3}$~$Z_\odot$, the mass in the halos is $100, 1000, 10^4$, and
$10^5$~M$_\odot$, respectively. More precise estimates can be obtained
using the metallicity distributions in \fig~\ref{fig:zmasf1} for each
simulation. Run S3, with strong feedback, gives smaller masses in
ultra low-metallicity stars than run S2, that has weak feedback.

Finally, we need to know how many stars have masses $M_* \simlt 1$
M$_\odot$, since more massive stars would evolve into unobservable
compact remnants. Assuming a Salpeter IMF, we estimate that $100$
M$_\odot$ in stars form about $100$ stars with masses $M_*<1$
M$_\odot$.  Therefore, the aforementioned values for the mass in stars
express also the number of ultra low metallicity stars, assuming a
Salpeter IMF.  If ultra low metallicity stars have a top-heavy IMF,
their number in the halo will be smaller. If there is a critical
metallicity that determines the transition from top-heavy to Salpeter
IMF, this should produced an observable feature in the observed number
counts of stars as a function of their metallicity, unless the
transition from top-heavy to Salpeter IMF is smooth.

\subsection{Internal Properties of the Galaxies}\label{ssec:int_prop}

In this section we study the internal structure of selected galaxies
in our highest resolution simulation (S1 in \tab~\ref{tab:one}) that,
in physical coordinates, has spatial resolution of about 10 pc at $z
\sim 9-10$. The stellar component of simulated dPri galaxies is a low
surface brightness spheroid that closely resembles dSph galaxies
observed in the Local Group.

\def\capfiggb{Azimuthally averaged gas (dots) and stellar (squares)
projected density profiles of six galaxies selected from the
simulation S1 at $z \sim 10$. For comparison, at virialization, the 
mean baryon density $\overline n_{vir} \sim 0.04$ cm$^{-3}$, and the 
density in the core is $n_{core} \sim 1$ cm$^{-3}$. The projected 
surface brightness, $\Sigma$ (dashed lines),
is shown on the right axis in units of magnitude per arcsec$^2$. The
numbers shown on the top right corners in each panel refer to the
totall mass in stars.} 
\placefig{
\begin{figure}[tb]
\epsscale{1.1}
\plotone{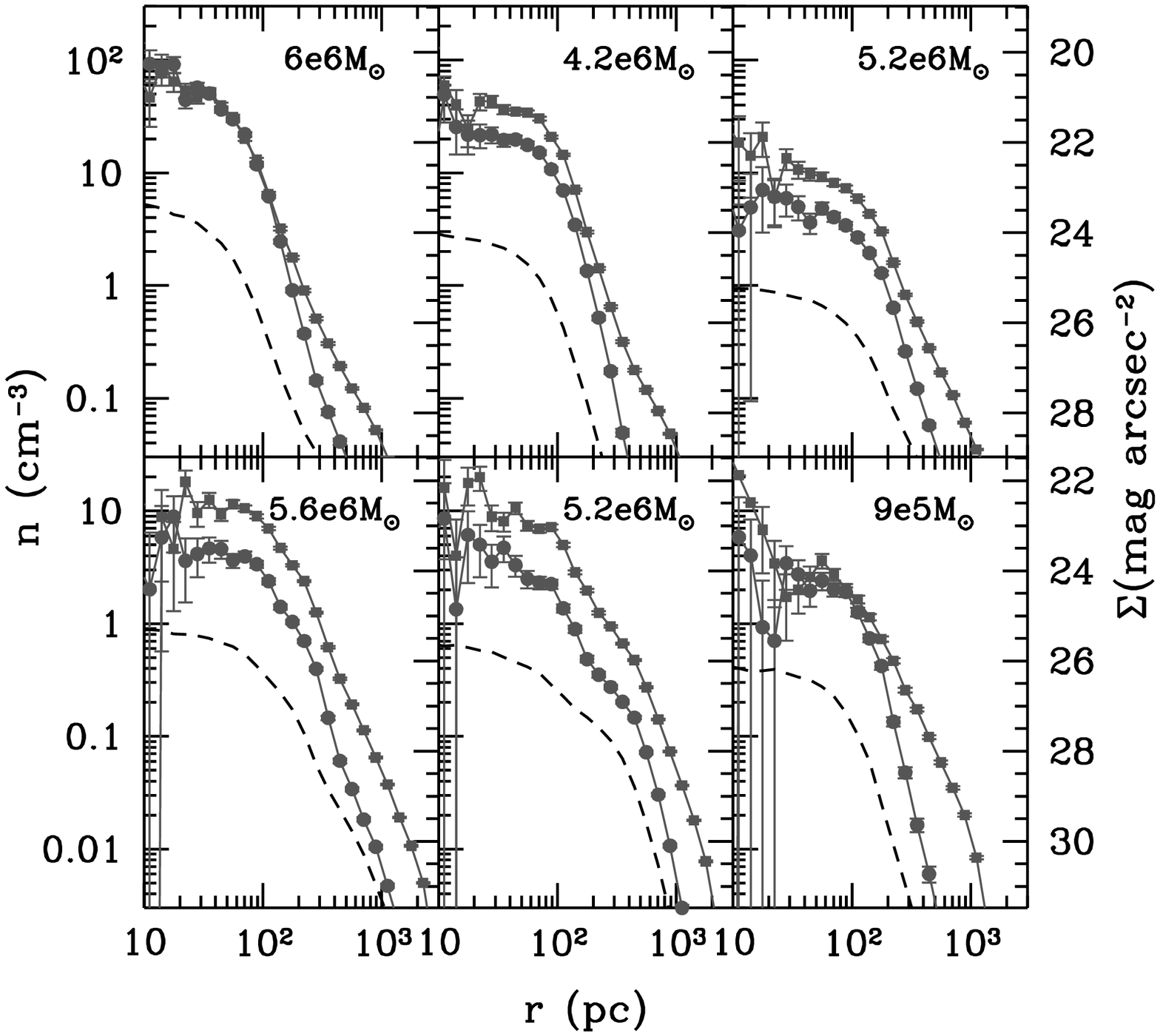}
\caption{\label{fig:prof}\capfiggb}
\end{figure}}
\def\capfiggc{Phase diagram of the ISM at metallicities $Z/Z_{\odot}$ 
  = 1, 0.01, and $10^{-3}$. The ISM in our simulated dPri galaxies have
  metallicities $Z = 0.01-0.1~Z_{\odot}$ and thermal pressures in the
  range shown by the cross-hatched band. Therefore, the conditions are
  compatible with a multi-phase ISM, where dense clouds and intercloud
  gas can coexist in a gas roughly at pressure equilibrium (at
  least in a broad statistical sense).} \placefig{
\begin{figure}[tb]
\epsscale{1.1}
\plotone{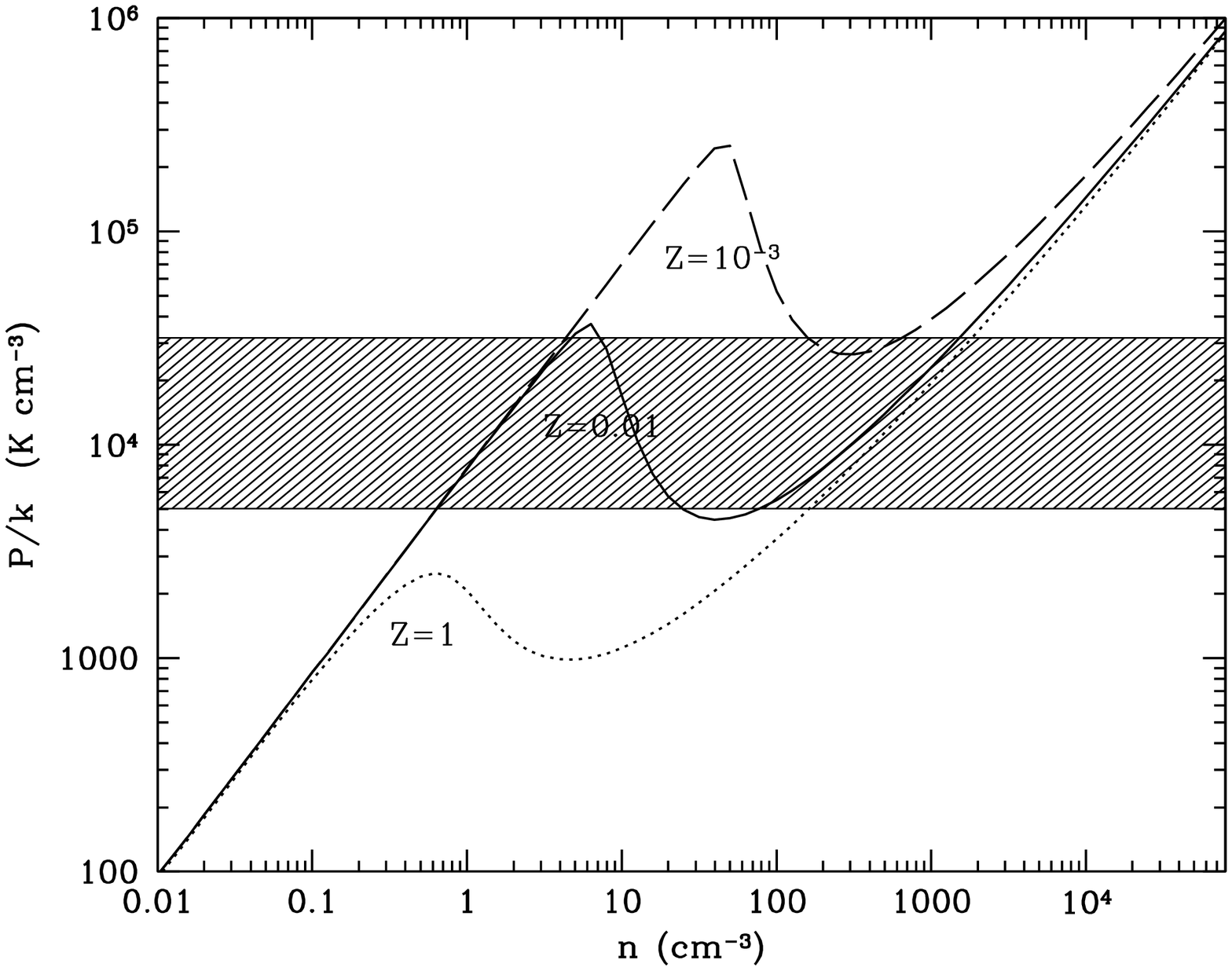}
\caption{\label{fig:phase}\capfiggc}
\end{figure}}
In \fig~\ref{fig:prof} we show the average density of gas (dots) and stars
(squares) assuming spherical shells, as a function of the distance
from the center of six galaxies selected from the simulation S1
at redshift $z \simeq 10$. The projected surface brightness, $\Sigma$
(dashed lines), is shown on the right axis.
\def\capfigha{
  ({\em Left}). Number density of primordial galaxies at redshifts
  $z=10.2, 12.5$, and $17.5$ as a function of their luminosity (or
  stellar mass) for the simulation S1. ({\em Right}). Star
  formation rate as a function of stellar mass for galaxies at
  redshifts $z=10.2, 12.5$, and $17.5$ in the simulation S1. The
  two lines that bracket most of the points show the star formation
  rate as a function of $M_*$ assuming that all the stars formed in a
  single burst of duration $t_{\rm burst} =10$ and $80$ Myr. }
\placefig{
\begin{figure*}[tb]
\epsscale{1.1}
\plottwo{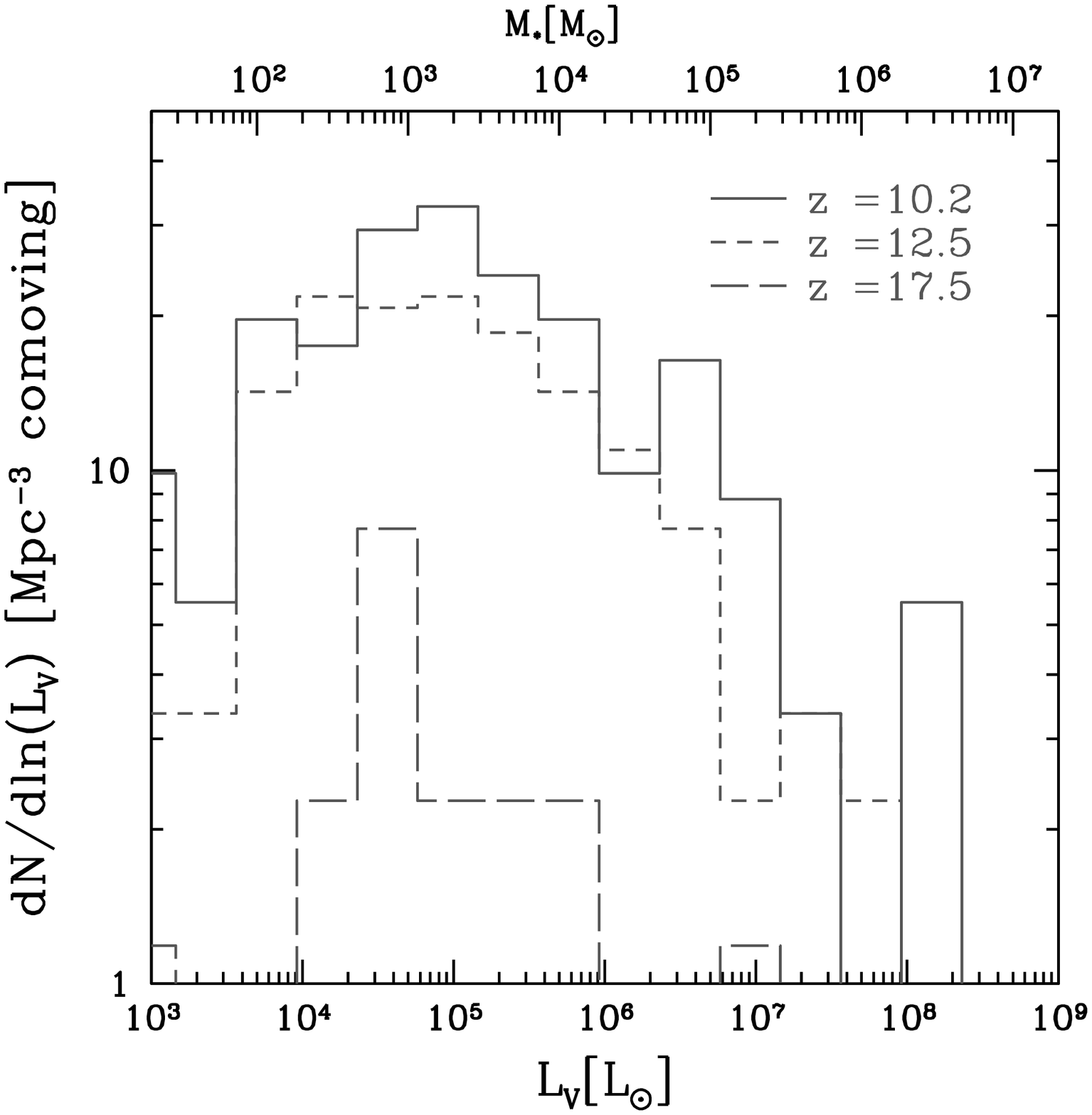}{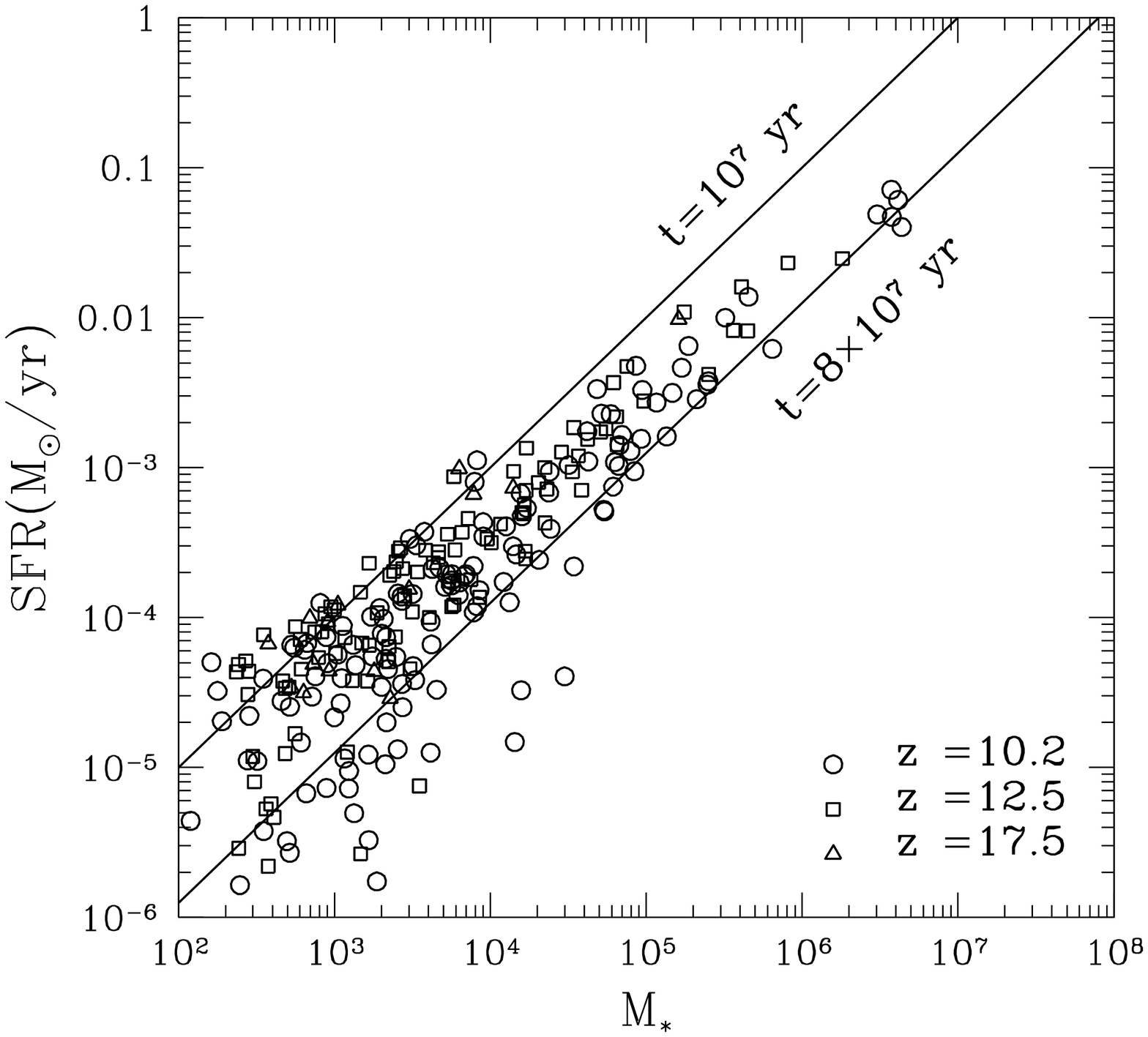}
\caption{\label{fig:NGST}\capfigha}
\end{figure*}} 

The gas density profile has a core of a few 100~pc in radius,
comparable to the core radius of the stellar component.  The mean gas
density in the core is $n_H \approx 10-50$ cm$^{-3}$.  The spatial
resolution of the simulation is not sufficient to resolve fluctuations
around this mean on scales smaller than 10-20 ~pc. It is therefore
impossible to study the structure of the multi-phase ISM. But two
interesting differences with respect to the Milky Way ISM structure
can be noted. The mean gas density in the core is 10--50 times
larger than the mean ISM density ($\sim 1$ cm$^{-3}$) in the Milky Way.
The gas and stellar component extend to the outer edges of the dark-matter 
halo.  The virial radii of the dark halos are about 1000 pc, only a few 
times more extended. For comparison, the Milky Way stellar spheroid has 
radius of a few kpc and the virial radius is about 300 kpc.  The mean 
temperature of the simulated ISM is $500-1000$~K, and the gas velocity 
dispersion is $\sigma_v \sim 10$ km s$^{-1}$, similar to the Milky-Way. 
The thermal pressure is therefore quite large, 
$P/k \sim 5 \times 10^3 - 5 \times 10^4$ cm$^{-3}$~K, several times larger 
than in the Milky Way \citep{JenkinsTripp:01,Wolfire:03}.  The ISM mean
metallicity is $Z \sim 0.1-0.01$ solar.  Assuming that the ISM is in
pressure equilibrium and including all relevant cooling and heating
processes as in \cite{Wolfire:95}, we have calculated the phase
diagram (thermal pressure as a function of density) for a gas of
metallicity $Z$ \citep[see][for details of the
calculation]{Ricotti:97}. The dotted, solid and dashed lines in
\fig~\ref{fig:phase} show the phase diagram of the ISM with
metallicity $Z/Z_{\odot} = 1$, 0.01, and $10^{-3}$, respectively. For
constant pressure, if the gas has metallicity $Z=0.01$ Z$_\odot$ and
the pressure is in the range shown by the shaded band, there are three
possible equilibrium values for the density: the low-density value is
thermally stable and is called warm neutral medium (WNM), the
intermediate density value is thermally unstable, and the high-density
value is stable and called cold neutral medium (CNM). If the pressure
lies outside this interval, a multi-phase medium in which dense clouds
(CNM) and inter-cloud gas (WNM) coexist cannot develop. From the
values of the pressure and metallicity in our simulated dPri galaxies,
we expect that a multi-phase ISM should be sustained.

We now return to the properties of the stellar component.  Most
simulated dark matter halos form stars very inefficiently or do not
form stars at all. The gas density profiles in these dark galaxies are
similar to the ones shown in \fig~\ref{fig:prof} and have a mean
molecular hydrogen abundance $x_{H_2} \sim 10^{-4}$. The high density
of the gas in these dark galaxies may have important implications for
reionization, through increases in the clumping of the IGM.  In
luminous galaxies the stellar component is a spheroid with negligible
angular momentum compared to the stellar velocity dispersion. The
spheroid has low surface brightness, but it is relatively extended,
with outer edges reaching $\sim 20$\% of the virial radius,
substantially more extended than in present-day galaxies.  The most
luminous galaxies in the simulation have surface brightness between
23-27 mag arcsec$^{-2}$, comparable to that of dSph galaxies in the
Local Group and Andromeda (\eg, 26.8 AndIX, 25.5 Ursa Minor, 25.3
Draco, 24.5 And~III). Detailed comparisons with the properties of dSph
galaxies was the subject of a separate paper \citep{RicottiG:05}. The
properties of the the dark matter halos were also discussed in
separate papers \citep{Ricotti:03, RicottiW:03} since they seem to
show cores instead of cusps, in agreement with many observations of
dwarf spheroidal galaxies and LSB galaxies \citep[\eg,][]{Kleyna:03,
Magorrian:03, deBlok:03}.

\section{Observability of Dwarf Primordial Galaxies}\label{sec:obs}

In this section, we study the feasibility of observing dPri galaxies
during their formation at redshifts $9<z<30$. In particular, we
calculate the number of point sources in the infrared bands detectable
in the field of view of the JWST. Then, we briefly address the
prospects for the identification of the fossil records of dPri
galaxies in the Local Group. Even if most dPri galaxies merge into
larger systems after their formation, a fraction of them
(approximatively $10$\%) is expected to survive to $z=0$. These relics
could be observed today as satellites of larger galaxies or in
isolation. We compare the mass function of the satellites observed
around the Milky Way and Andromeda with the simulated mass function of
luminous halos. For realistic values of the survival probablity, our
simulations, which include radiative feedback but not the effect of
reionization, can reproduce the observed mass function of Galactic
satellites.

\subsection{Detection of Primordial Galaxies with JWST}

The luminosity function of dPri galaxies at redshifts $z=10.2, 12.5$,
and $17.5$ is shown in the \fig~\ref{fig:NGST}~(left) for simulation S1. 
We assumed a mass-to-light ratio $M_*/L_V =1/50$ (solar),
appropriate for starbursts with age $t \sim 100$ Myr. The panel on the
right shows the star formation rate as a function of the stellar mass
for the same galaxies. The two lines show the values of the mean star
formation rate (SFR) that would produce a stellar mass, $M_*$, in a
single burst of durations $t_{\rm burst} = M_*/SFR \simeq 10$ Myr and 
$80$ Myr.

About 10\% of dwarf sized dark matter halos with $M_{\rm dm}>10^6$
M$_\odot$ that assembled prior to reionization are able to form stars.
There are $\sim 500$ dPri galaxies per Mpc$^{3}$ with luminosities
spanning four decades, between $10^4$ and $10^8$ L$_\odot$. The
luminosity function is rather flat, with $\sim10$ galaxies Mpc$^{-3}$
in the higher luminosity decade ($10^7<L<10^8$ L$_\odot$) and $200$
Mpc$^{-3}$ in the fainter decade ($10^4<L<10^5$ L$_\odot$).

The integrated number counts of galaxies at $z>9$ in the IR bands is
shown in \fig~\ref{fig:Ncount}. 
The solid lines show the number counts at $z<$ and $z<$ derived from
the simulation data. The dashed lines are extrapolations of the bright
end of the luminosity function to account for more massive galaxies
not present in our simulations due to their small volume. The dashed
lines are calculated using Press-Schechter formalism to derive the
number counts of dark halos and assuming a stellar mass in each of
them that is a constant fraction of their total mass (hence neglecting
feedback). The assumed constant value of the star formation efficiency
is the same as for the most massive halos present in the
simulation. Although the counts are uncertain, especially at the
bright end of the luminosity function, \fig~\ref{fig:Ncount} clearly
shows that JWST will not be able to detect any of the small mass
galaxies formed in our simulations.

For a Salpeter IMF and $\langle f_{esc}\rangle \sim 1$, JWST might be
able to detect the most massive galaxies at redshift $z \sim 10$. But
the prospects for observing the formation of the fainter low-mass
galaxies that cool by H$_2$ is very small. The number of faint sources
will not be able to resolve the controversy of whether dark halos with
$M_{\rm }<10^8$ M$_\odot$ host luminous galaxies, or have their
formation suppressed by H$_2$-dissociating radiation. But it may be
possible to infer the slope of the luminosity function below the
sensitivity limit of the JWST by analyzing the fluctuation of the
unresolved background \citep[see for example,][for a technique applied
to the {\it Chandra} deep field]{Miyaji:02}.  Only if the IMF in dPri
galaxies is top-heavy and $\langle f_{esc}\rangle\ll 1$ the luminosity
of dPri galaxies in the K bands (rest frame UV) at the same mass will
be up to 10 times higher than in \fig~\ref{fig:Ncount}. In this second
case, it may be possible to use the faint number counts to constrain
the theoretical models and quantify the relative importance of
negative and positive feedback on the formation of dPri galaxies.

\def\capfighb{Cumulative number counts of primordial galaxies in the K
  bands at redshifts $z \simgt 9$ in our higher resolution simulation
  (S1). Similar results are obtained for the other
  simulations. Assuming a Salpeter (top-heavy) IMF, dwarf primordial
  galaxies with masses $M_{\rm dm}>10^8$ M$_\odot$ ($M_{\rm dm}>10^7$
  M$_\odot$) are luminous enough to be detectable with the JWST, if it
  operates at nominal sensitivity shown by the vertical line (NIRCAM
  has a sensitivity of 3~nJy in the F200W and F277W filters for $10^5$
  sec exposures).}  \placefig{
\begin{figure}[tb]
\epsscale{1.0}
\plotone{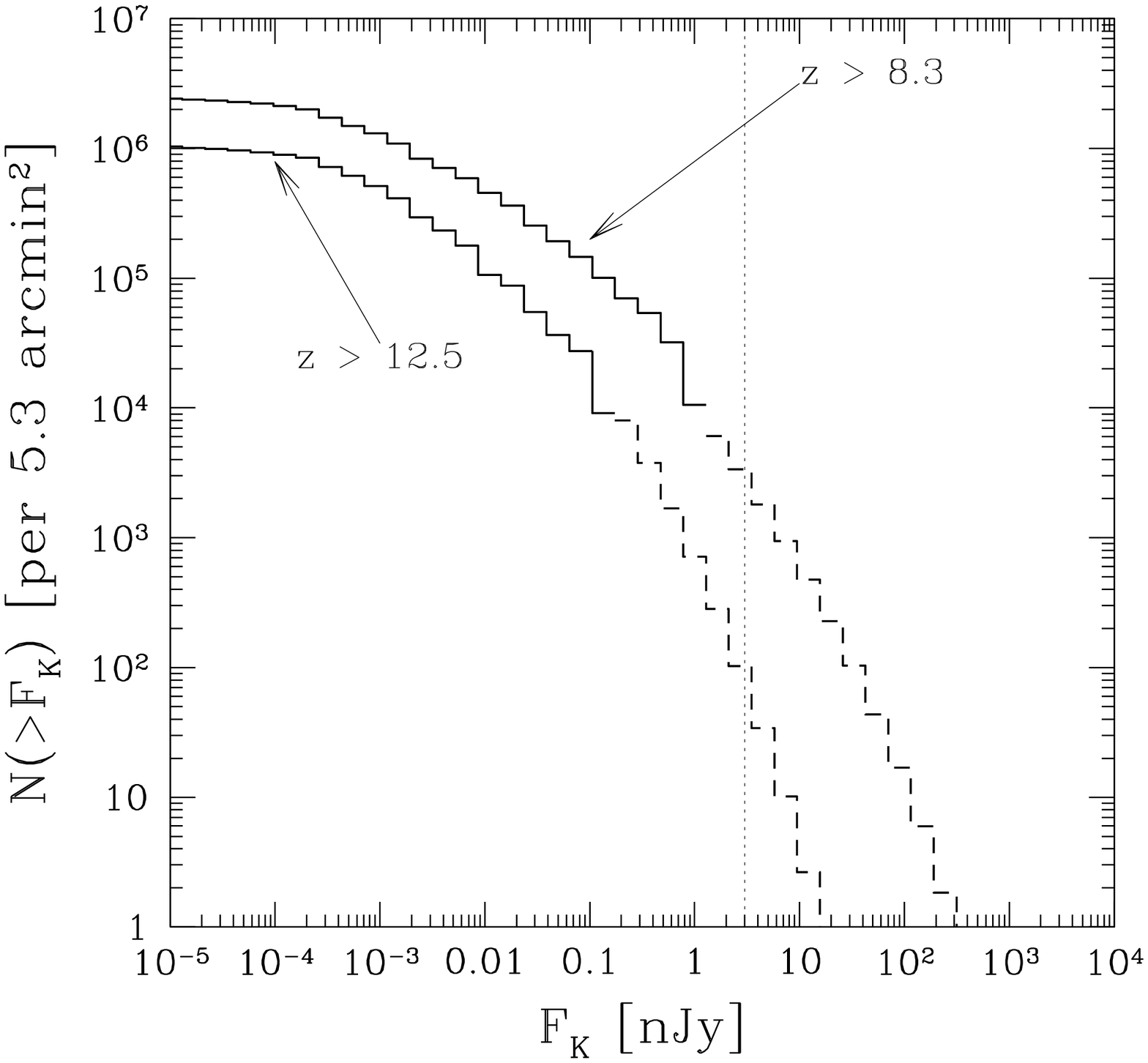}
\caption{\label{fig:Ncount}\capfighb}
\end{figure}
}
\subsection{Relics of Primordial Dwarf Galaxies}\label{subsec:relic}

How many galaxies that formed the bulk of their stars before
reionization, do we expect to observe in the Local Universe? In CDM
cosmologies, galaxies similar to the Milky Way were formed by
accreting the debris of old, lower-mass galaxies and the intergalactic
gas surrounding them. However, most low-mass galaxies that were the
dominant galaxy population at high redshift, have been destroyed and
incorporated into larger galaxies, constituting a fraction of their
bulge and halo stars.  The probability that a galaxy formed at
redshift $z_f$ survives without being incorporated into a larger one
is roughly $(1+z_f)^{-1}$ \citep[\eg,][]{Sasaki:94}.  Therefore, about
10\% of the galaxies in our simulations are expected to survive to the
present.  Since the total number of low-mass galaxies per
$h^{-3}$~Mpc$^{3}$ at $z=10$ is 10--100, the number of fossil
primordial galaxies today should be about $1-10$ per Mpc$^3$ $h^{-3}$.
\def\capfighc{Mass function of dark matter halos (histograms) and
  luminous galaxies (shaded histograms) at $z \approx 10$ in four
  simulations from \tab~\ref{tab:one}. The left panels show run S1
  (high-resolution run) and run S2. The right panels show run S3
  (strong radiative feedback) and S4 (without radiative feedback). The
  shaded histograms in the two top panels display the mass function of
  dwarf galaxies with $M_* \ge 5 \times 10^5$ M$_\odot$ (or $L_V
  \simgt 10^5$ L$_\odot$ assuming an old stellar population) and the
  bottom panels with $M_* \ge 5 \times 10^3$ M$_\odot$ (or $L_V \simgt
  10^3$ L$_\odot$), thus including ultra-faint dwarfs.}  \placefig{
\begin{figure*}[tb]
\epsscale{1.0}
\plotone{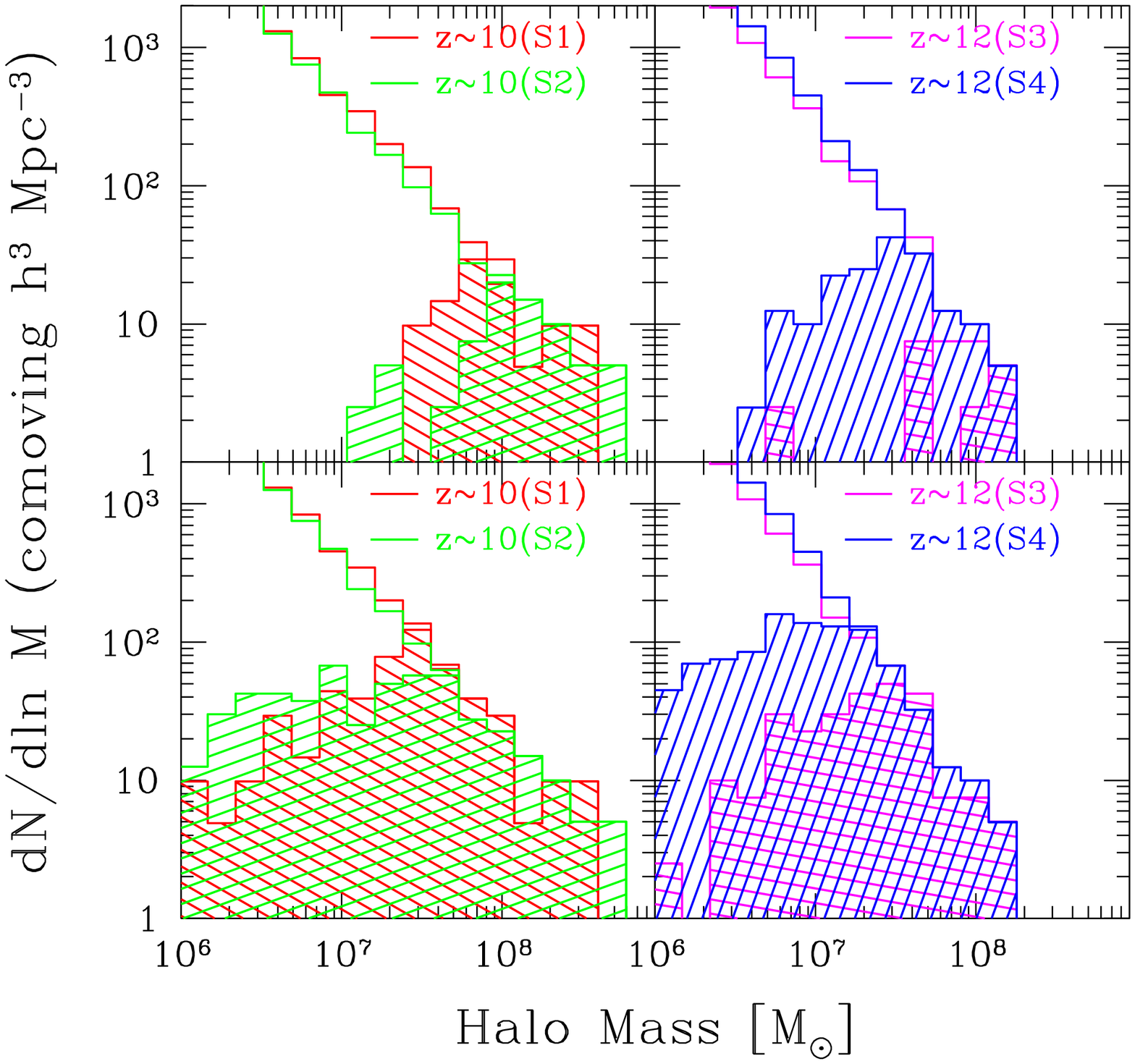}
\caption{\label{fig:massf}\capfighc}
\end{figure*}
} 

A detailed study on the identification of the fossils of the first
galaxies in the Local Group is the focus of two previous works
\citep{RicottiG:05, GnedinK:06}. In \citep[][, hereafter
RG05]{RicottiG:05}, we have focused on comparing observable properties
of the population of primordial dwarf galaxies from our
high-resolution simulation (run S1), to the properties of Local Group
dwarf galaxies. In RG05 we have evolved the run S1 to lower redshift
and have introduced a strong source of ionization in the simulated
volume that fully reionize the IGM (see the paper for details). After
reionization, due to the increase of the IGM Jeans mass, low-mass
halos are devoided of all gas and are not able to form new stars
because gas accretion is suppressed by IGM reheating. Hence, the
subset of galaxies in our simulation that survives tidal stripping,
can be compared to present day galaxies simply by aging passively
their stellar populations. The striking similarities between the
properties simulated galaxies and many Local Group dSphs lead us to
propose a scenario for their origin as the surviving well-preserved
fossils of the first galaxies. In addition, the simulation shows the
existence of a population of ultra-faint dwarf galaxies, not observed
at the time of publication. This ultra-faint population may have been
recently discovered in the Local Group
\citep{Belokurovetal07,Belokurovetal06a,Irwinetal07,Willmanetal05ApJ,
Willmanetal05AJ,Walshetal07,Zuckeretal06a,Zuckeretal06b,Ibataetal07,
Majewskietal07,Martinetal06}.

\cite{GnedinK:06}, presented a detailed study of the prediction of
the simulation in RG05, on the probability of survival of the fossils
of dPri galaxies as they are incorporated in a Milky Way type
halo. The results of this study show that the galactocentric
distribution of the simulated galaxies reproduce the observed
distribution of normal dwarf around the Milky-Way, but the ultra-faint
population of simulated dwarfs was not accounted for at the time of
publication.  Finally, in \citet{BovillR:08} we show that the
properties of the newly discovered population of dwarf galaxies matches
the theoretical prediction in RG05 and \citep{GnedinK:06}. The
galactocentric distribution of the recently discovered population of
ultra-faint dwarfs nearly closes the gap between observations and
theoretical predictions, hence solving the well know ``missing
galactic satellite problem''.

In this section we focus on another well know observed property of the
normal population (as opposed to the ultra-faint one, with $L_V <
10^5~L_\odot$ or $M_V \simlt -7.7$) of dwarf galaxies in the Local
Group: there seems to be a characteristic dynamical mass of about
$10^7$~M$_\odot$.  One may be tempted to assume that this mass is the
smallest galactic halo mass, or the galaxy building block. However,
some of the newly discovered ultra-faint dwarfs have dynamical masses
that are smaller than in the normal population \citep{SimonG:07}.
Here, comparing different simulations in \tab~\ref{tab:one}, we show
that radiative feedback processes determine the value of the
characteristic mass of $10^7$ M$_\odot$ in the normal population of
dSphs, and that the strength of radiative feedback determines the
number and characteristic mass (typically $<10^7$ M$_\odot$), of the
ultra-faint dwarf population.  In order to fully understand the origin
and the physics of the mass cut-off, much more work is needed. This
study goes beyond the scope of the present paper and will be explored
separately.

The shaded histograms in \fig~\ref{fig:massf} show the mass function
of galaxies at $z \approx 10$ (\ie, luminous halos) compared to the
mass function of all halos (\ie, dark and luminous) in four
simulations from \tab~\ref{tab:one}.  The shaded histograms in the top
panels show the mass function of ``normal'' dwarf galaxies with $L_V
\ge 10^5$ L$_\odot$ ($M_V < -7.7$), while the two bottom panels
include also ultra-faint dwarfs with $L_V \ge 10^3$ L$_\odot$.  Run S1
(the high resolution run) and run S2 (the run with weak radiative
feedback) are shown in the left panels. Run S3 with strong feedback,
and run S4 withouth radiative feedback are shown in the righ panels.
The panels illustrate two important properties of the first
galaxies:\\ 1) Dwarfs that have luminosities comparable or larger than
Ursa Minor or Draco (top panels), have a mass function with a
characteristic mass cut off $M_{dm} \sim 10^7$~M$_\odot$, in agreement
with observations \citep[\eg,][]{Mateo:98}.  The mass cut off is not a
numerical artifact - it is lower in the lower resolution runs S2 and
S3 than in the higher resolution run, S1.  Clearly, the mass cutoff is
produced by radiative feedback, as it is not present in the simulation
without radiative feedback, S4, and it is weakly dependent on the
strength of feedback (\eg, compare run S2 and S3).\\ 2) The
simulations show the existence of a population of ultra-faint dwarfs
with $L_V<10^{5}$ L$_\odot$ (bottom panels). The mass function of this
ultra-faint population extends down to the smallest masses resolved in
our simulations, or shows a smaller lower mass cut off than in normal
dwarfs. If feedback is strong as in run S2, the ultra faint galaxies
are fewer and have larger typical masses than in the runs with weak or
no radiative feedback.

Here, it is important to recall our definition of strong and weak
feedback: whether a simulations has weak or strong radiative feedback,
depends on the intensity of ultraviolet radiation that escapes into
the IGM. Hence, a run with top-heavy IMF and \fesc$\sim 1$ is a
simulation with strong feedback, while a run with Salpeter IMF and
\fesc$< 1$ corresponds to a simulation with weak feedback.

\section{Summary and Conclusions}\label{sec:sum}

In this paper, the third of a series, we have analyzed in detail the
properties of the first galaxies in a set of three-dimensional
cosmological simulations.  This paper is devoted to the analysis of 
statistical and internal properties of the population of dPri galaxies 
and their impact on the IGM.  

In the first part (Paper~1) of this study,
we addressed the problem of simulating the formation of the first
galaxies by implementing a cosmological code that includes
time-dependent, three-dimensional radiative transfer of \HI, \GI and
\GII ionizing photons in a cosmological volume. Using a recipe for
star formation and a fast method to solve radiative transfer
\citep{GnedinA:01} at each hydro time-step, we were able to simulate
the radiative feedback processes by the first luminous sources. We
modelled the physics of chemically pristine gas including a
non-equilibrium treatment of the chemistry of nine species (e$^-$, H,
H$^+$, He, He$^+$, He$^{+2}$, H$_2$, H$^+_2$, H$^-$), cooling by
molecular hydrogen, ionization by secondary electrons, Ly$\alpha$
pumping, and radiative transfer for the narrow lines in the H$_2$
Lyman-Werner bands of the dissociating background radiation.
Currently, these simulations are the state of the art for the formation
of dPri
galaxies, in the sense that are the only simulations that include
time-dependent and spatially inhomogeneous radiative feedback
processes without introducing sub-grid analytical recipes.

In the second part of this study (Paper~2) we found that, contrary to
previous work, dPri galaxies are able to form enough stars to be
cosmologically important.  The reasons for disagreement of our results
with previous ones may be due to the density-dependent reformation
rate of H$_2$ (quickly photodissociated in the voids, but not in the
denser filaments) and time-dependent feedback produced by the bursting
mode of star formation (\eg, H$_2$ reformation in relic \HII regions)
or a combination of both. These processes are not included in previous
semi-analytic studies and numerical simulations \citep{Machacek:00,
Tassis:03}.  Our simulations do not include the effect of H$_2$
self-shielding by photodissociating radiation, so it is possible that
we are underestimating star formation in lower-mass
galaxies. However, we have shown that the global star formation rate
is self-regulated and insensitive to the intensity of the
photodissociating background. In most simulations (see
\tab~\ref{tab:one}) we did not include mechanical feedback by SN
explosions, motivated by previous results \citep{Gnedin:98a} showing
that their effect is negligible unless we adopt a top-heavy IMF. The
effect of SN explosions depends strongly on the particular
implementation in the code and is difficult to test.  The method for
solving radiative transfer has instead been tested on simple benchmark
problems, but a better test of the reliability of our results will
need to wait for other simulations that include a similar
self-consistent treatment of radiative transfer and feedback.

In the following list we summarize the main conclusions of our
simulations of properties of our sample of dPri galaxies and their
effects on the metallicity of the IGM.
\begin{enumerate}

\item {\em Number and luminosity of dPri galaxies}. About 10\% of
  dwarf dark matter halos ($M_{\rm dm}>10^6$ M$_\odot$) assembled prior to 
  reionization are able to form stars. We find $\sim 500$ dPri galaxies per
  Mpc$^{3}$ between $10^4$ and $10^8$ L$_\odot$. The luminosity
  function is rather flat, with $10$ galaxies Mpc$^{-3}$ 
  ($10^7<L<10^8$ L$_\odot$) and $200$ Mpc$^{-3}$ at ($10^4<L<10^5$ L$_\odot$).

\item {\em Relative importance of H$_2$ and metal cooling.} H$_2$
  cooling is important for the formation of the first few stars in
  each protogalaxy. As the first few stars form, if the ISM has not
  been blown out, Ly$\alpha$ and metal-line cooling become dominant.
  If radiative feedback is strong (top-heavy IMF)
  star formation in lower mass dark halos is suppressed. If the
  feedback is weak (Salpeter IMF and/or small $\langle f_{esc}\rangle$), 
  star formation in lower-mass galaxies is inefficient and delayed, 
  but not suppressed.

\item {\em Clustering.} The local nature of feedback has implications 
  for the clustering and bias properties of the first luminous galaxies.
  Analogous to young star clusters at low-redshift, these form 
  preferentially in groups and chain-like structures and are more 
  clustered than the dark halos of the same mass.

\item {\em Volume filling factor of metal-enriched IGM}.  The metals
  produced by the first galaxies can fill the space between bright
  galaxies rather uniformly, but only to very low values of the
  metallicity ($Z/Z_\odot \simlt 10^{-5}$).  The volume filling factor
  of the IGM enriched to the typical metallicities observed in the
  Ly$\alpha$ forest is small. It is unlikely that the metal absorption systems 
  seen in the Ly$\alpha$ forest were produced by the first low-mass galaxies.

\item {\em Gas photoevaporation.}  Star-forming dwarf galaxies show
  large variations in their gas content because of the combined
  effects of stellar feedback from internal sources and IGM reheating.
  Ratios of gas to dark matter lie below the cosmic mean in halos with
  masses $M_{\rm dm}<10^8$ M$_\odot$. Gas depletion increases with
  decreasing redshift: the lower-mass halos lose all their gas first
  but, as the universe evolves, larger halos with $M_{\rm dm} \sim
  10^8$ M$_\odot$ also lose a large fraction of their gas.

\item {\em Mean star-formation efficiency}. The mean star formation
  efficiency $\langle f_*(t) \rangle = \langle M_*/M_{\rm bar}^{\rm max}
  \rangle$. We assume that $M_{\rm bar}^{\rm max} \simeq M_{\rm dm}/7$,
  with an efficiency independent of redshift and depending on total mass 
  as $\langle f_* (t) \rangle \propto M_{\rm dm}^{\alpha}$. There is 
  weak dependence on feedback: $\alpha=1.5$ if the feedback is weak 
  and $\alpha=2$ if the feedback is strong.

\item {\em Scatter of the mass-to-light ratio}. A tight relationship
  between the star formation efficiency $f_*$ and the total mass of
  halos holds only for galaxies with $M_{\rm dm}>5 \times 10^7$ M$_\odot$.
  In lower-mass halos, the scatter around the mean $\langle f_*
  \rangle$ is increasingly large. For a given halo mass, the galaxy
  can be without stars (dark galaxy, $f_*=0$) or have $f_* \sim 0.5$.
  A few dark galaxies as massive as $M_{\rm dm} \sim 1-5 \times
  10^{7}$ M$_\odot$ may exist in the Local Group.

\item {\em Low-metallicity stars}. The mass fraction of \pop3\ stars
  (metallicity $Z<10^{-3}$) with respect to \popII\ stars is about
  one in a million at $z=10$. The epoch dominated by \pop3\ stars 
  is short-lived.  In models with strong feedback, it ends at $z \sim 17$,
  while if the feedback is weak, a small fraction of \pop3\ stars is still
  forming at $z=10$. About $N=1000 (Z_0/10^{-3} Z_\odot)$ stars with
  metallicity smaller than a floor $Z_0$ should be present in the Galactic
  halo for a Salpeter IMF. If the feedback is strong, the
  distribution of low-metallicity stars has a sharp drop at $Z \simeq
  10^{-4}$ Z$_\odot$ and is almost flat at $10^{-4} < Z/Z_\odot <
  10^{-2}$. Furthermore, if there is a transition from
  a Salpeter IMF to top-heavy IMF at a critical value of the stellar
  metallicity ($Z<Z_{\rm cr}$), the distribution should show a cutoff
  at that critical metallicity.

\item {\em Luminosity profile}. Galaxies with masses $M_{\rm }<10^8$
  M$_\odot$ have a low surface brightness and extended stellar spheroid.  
  The outer edges of the stellar spheroid nearly reaches the virial 
  radius. In more massive galaxies that cool efficiently by Ly$\alpha$ 
  radiation, the stars and light are more centrally concentrated.

\item {\em Multi-phase interstellar medium.} The interstellar medium
  (ISM) of these galaxies has mean density $10-100 ~{\rm cm}^{-3}$ and 
  thermal pressure $P/k \sim 10^5 ~{\rm cm}^{-3}$~K. This pressure is
  sufficiently high to develop a multi-phase ISM in low
  metallicity gas ($10^{-2}-10^{-3}~Z_\odot$).

\item {\em Deep field number counts}.  For a Salpeter IMF,
  JWST might detect only the most massive dPri galaxies. The prospects
  for observing the formation of the fainter low-mass galaxies that
  cool by H$_2$ is very small.  If, instead, the IMF in
  dPri galaxies is top-heavy and $\langle f_{esc}\rangle \ll 1$, JWST
  might be able to detect dPri galaxies with masses as small as
  $M_{\rm dm} \sim 10^7$ M$_\odot$. In this second case, the faint
  number counts could constrain theoretical models and quantify the
  relative importance of negative and positive feedback for their
  formation.

\item {\em Typical masses of dSphs and ultra-faint dwarfs in the Local
  Group}. Radiative feedback suppresses star formation in low-mass
  halos.  Halos with $M_{dm} > 10^7$ M$_\odot$ produce galaxies with
  luminosities typical of normal dSphs such as Ursa Minor and
  Draco. Halos with $M_{dm} < 10^7$ M$_\odot$ may host a population
  of ultra-faint dwarfs, similar to the one recently discovered in the
  Local Group, or may be completely dark.
\end{enumerate}

\acknowledgements 

This work was supported by the Theoretical Astrophysics program at the
University of Colorado (NASA grant NNX07AG77G and NSF grant AST07-07474) 
and at the University of Maryland (NASA grant NNX07AH10G).
The simulations were performed using SGI/CRAY Origin 2000 array at the
National Center for Supercomputing Applications (NCSA).

\clearpage
\bibliographystyle{/home/ricotti/Latex/TeX/apj}
\bibliography{/home/ricotti/Latex/TeX/archive}

\begin{thebibliography}{}

\bibitem[\protect\citeauthoryear{{Abel}, {Bryan}, \& {Norman}}{{Abel}
  et~al.}{2002}]{Abel:02}
{Abel}, T., {Bryan}, G.~L.,  \& {Norman}, M.~L. 2002, Science, 295, 93

\bibitem[\protect\citeauthoryear{{Adelberger} et~al.}{{Adelberger}
  et~al.}{2003}]{Adelberger:03}
{Adelberger}, K.~L., {Steidel}, C.~C., {Shapley}, A.~E.,  \& {Pettini}, M.
  2003, \apj, 584, 45

\bibitem[\protect\citeauthoryear{{Ahn} \& {Shapiro}}{{Ahn} \&
  {Shapiro}}{2007}]{Ahn:07}
{Ahn}, K.,  \& {Shapiro}, P.~R. 2007, \mnras, 375, 881

\bibitem[\protect\citeauthoryear{{Belokurov} et~al.}{{Belokurov}
  et~al.}{2007}]{Belokurovetal07}
{Belokurov}, V., et~al. 2007, \apj, 654, 897

\bibitem[\protect\citeauthoryear{{Belokurov} et~al.}{{Belokurov}
  et~al.}{2006}]{Belokurovetal06a}
{Belokurov}, V., et~al. 2006, \apjl, 647, L111

\bibitem[\protect\citeauthoryear{{Bovill} \& {Ricotti}}{{Bovill} \&
  {Ricotti}}{2008}]{BovillR:08}
{Bovill}, M.,  \& {Ricotti}, M. 2008, in preparation

\bibitem[\protect\citeauthoryear{{Bovill} et~al.}{{Bovill}
  et~al.}{2008}]{Bovill:08}
{Bovill}, M., {Ricotti}, M., {Gnedin}, Y.,  \& {Kravstov}, A. 2008, in
  preparation

\bibitem[\protect\citeauthoryear{{Bromm}, {Coppi}, \& {Larson}}{{Bromm}
  et~al.}{1999}]{BrommCL:99}
{Bromm}, V., {Coppi}, P.~S.,  \& {Larson}, R.~B. 1999, \apjl, 527, L5

\bibitem[\protect\citeauthoryear{{Couchman} \& {Rees}}{{Couchman} \&
  {Rees}}{1986}]{CouchmanR:86}
{Couchman}, H.~M.~P.,  \& {Rees}, M.~J. 1986, \mnras, 221, 53

\bibitem[\protect\citeauthoryear{{de Blok}, {Bosma}, \& {McGaugh}}{{de Blok}
  et~al.}{2003}]{deBlok:03}
{de Blok}, W.~J.~G., {Bosma}, A.,  \& {McGaugh}, S. 2003, \mnras, 340, 657

\bibitem[\protect\citeauthoryear{{Ferrara}}{{Ferrara}}{1998}]{Ferrara:98}
{Ferrara}, A. 1998, \apjl, 499, L17

\bibitem[\protect\citeauthoryear{{Ferrara}, {Pettini}, \&
  {Shchekinov}}{{Ferrara} et~al.}{2000}]{Ferrara:00}
{Ferrara}, A., {Pettini}, M.,  \& {Shchekinov}, Y. 2000, \mnras, 319, 539

\bibitem[\protect\citeauthoryear{{Fujita} et~al.}{{Fujita}
  et~al.}{2004}]{Fujita:04}
{Fujita}, A., {Mac Low}, M.-M., {Ferrara}, A.,  \& {Meiksin}, A. 2004, \apj,
  613, 159

\bibitem[\protect\citeauthoryear{{Gnedin}}{{Gnedin}}{1998}]{Gnedin:98a}
{Gnedin}, N.~Y. 1998, \mnras, 294, 407

\bibitem[\protect\citeauthoryear{{Gnedin} \& {Abel}}{{Gnedin} \&
  {Abel}}{2001}]{GnedinA:01}
{Gnedin}, N.~Y.,  \& {Abel}, T. 2001, New Astronomy, 6, 437

\bibitem[\protect\citeauthoryear{{Gnedin} \& {Kravtsov}}{{Gnedin} \&
  {Kravtsov}}{2006}]{GnedinK:06}
{Gnedin}, N.~Y.,  \& {Kravtsov}, A.~V. 2006, \apj, 645, 1054

\bibitem[\protect\citeauthoryear{{Haiman}, {Abel}, \& {Rees}}{{Haiman}
  et~al.}{2000}]{HaimanAR:00}
{Haiman}, Z., {Abel}, T.,  \& {Rees}, M.~J. 2000, \apj, 534, 11

\bibitem[\protect\citeauthoryear{{Haiman} \& {Bryan}}{{Haiman} \&
  {Bryan}}{2006}]{HaimanB:06}
{Haiman}, Z.,  \& {Bryan}, G.~L. 2006, \apj, 650, 7

\bibitem[\protect\citeauthoryear{{Haiman}, {Rees}, \& {Loeb}}{{Haiman}
  et~al.}{1996}]{HaimanRL:96}
{Haiman}, Z., {Rees}, M.~J.,  \& {Loeb}, A. 1996, \apj, 467, 522

\bibitem[\protect\citeauthoryear{{Ibata} et~al.}{{Ibata}
  et~al.}{2007}]{Ibataetal07}
{Ibata}, R., {Martin}, N.~F., {Irwin}, M., {Chapman}, S., {Ferguson}, A.~M.~N.,
  {Lewis}, G.~F.,  \& {McConnachie}, A.~W. 2007, \apj, 671, 1591

\bibitem[\protect\citeauthoryear{{Irwin} et~al.}{{Irwin}
  et~al.}{2007}]{Irwinetal07}
{Irwin}, M.~J., et~al. 2007, \apjl, 656, L13

\bibitem[\protect\citeauthoryear{{Jenkins} \& {Tripp}}{{Jenkins} \&
  {Tripp}}{2001}]{JenkinsTripp:01}
{Jenkins}, E.~B.,  \& {Tripp}, T.~M. 2001, \apjs, 137, 297

\bibitem[\protect\citeauthoryear{{Johnson} \& {Bromm}}{{Johnson} \&
  {Bromm}}{2007}]{JohnsonB:07}
{Johnson}, J.~L.,  \& {Bromm}, V. 2007, \mnras, 374, 1557

\bibitem[\protect\citeauthoryear{{Kleyna} et~al.}{{Kleyna}
  et~al.}{2003}]{Kleyna:03}
{Kleyna}, J.~T., {Wilkinson}, M.~I., {Gilmore}, G.,  \& {Evans}, N.~W. 2003,
  \apjl, 588, L21

\bibitem[\protect\citeauthoryear{{Lavery} \& {Mighell}}{{Lavery} \&
  {Mighell}}{1992}]{Lavery:92}
{Lavery}, R.~J.,  \& {Mighell}, K.~J. 1992, \aj, 103, 81

\bibitem[\protect\citeauthoryear{{Machacek}, {Bryan}, \& {Abel}}{{Machacek}
  et~al.}{2001}]{Machacek:00}
{Machacek}, M.~E., {Bryan}, G.~L.,  \& {Abel}, T. 2001, \apj, 548, 509

\bibitem[\protect\citeauthoryear{{Machacek}, {Bryan}, \& {Abel}}{{Machacek}
  et~al.}{2003}]{Machacek:03}
{Machacek}, M.~E., {Bryan}, G.~L.,  \& {Abel}, T. 2003, \mnras, 338, 273

\bibitem[\protect\citeauthoryear{{Madau}, {Ferrara}, \& {Rees}}{{Madau}
  et~al.}{2001}]{MadauF:01}
{Madau}, P., {Ferrara}, A.,  \& {Rees}, M.~J. 2001, \apj, 555, 92

\bibitem[\protect\citeauthoryear{{Madau}, {Meiksin}, \& {Rees}}{{Madau}
  et~al.}{1997}]{Madau:97}
{Madau}, P., {Meiksin}, A.,  \& {Rees}, M.~J. 1997, \apj, 475, 429

\bibitem[\protect\citeauthoryear{{Madau} et~al.}{{Madau}
  et~al.}{2004}]{MadauR:03}
{Madau}, P., {Rees}, M.~J., {Volonteri}, M., {Haardt}, F.,  \& {Oh}, S.~P.
  2004, \apj, 604, 484

\bibitem[\protect\citeauthoryear{{Magorrian}}{{Magorrian}}{2003}]{Magorrian:03}
{Magorrian}, J. 2003, in The Mass of Galaxies at Low and High Redshift.
  Proceedings of the ESO Workshop held in Venice, Italy, 24-26 October 2001, p.
  18.

\bibitem[\protect\citeauthoryear{{Majewski} et~al.}{{Majewski}
  et~al.}{2007}]{Majewskietal07}
{Majewski}, S.~R., et~al. 2007, \apjl, 670, L9

\bibitem[\protect\citeauthoryear{{Martin} et~al.}{{Martin}
  et~al.}{2006}]{Martinetal06}
{Martin}, N.~F., {Ibata}, R.~A., {Irwin}, M.~J., {Chapman}, S., {Lewis}, G.~F.,
  {Ferguson}, A.~M.~N., {Tanvir}, N.,  \& {McConnachie}, A.~W. 2006, \mnras,
  371, 1983

\bibitem[\protect\citeauthoryear{{Mateo}}{{Mateo}}{1998}]{Mateo:98}
{Mateo}, M.~L. 1998, \araa, 36, 435

\bibitem[\protect\citeauthoryear{{Miller} et~al.}{{Miller}
  et~al.}{2003}]{Miller:03}
{Miller}, J.~M., {Fabbiano}, G., {Miller}, M.~C.,  \& {Fabian}, A.~C. 2003,
  \apjl, 585, L37

\bibitem[\protect\citeauthoryear{{Miyaji} \& {Griffiths}}{{Miyaji} \&
  {Griffiths}}{2002}]{Miyaji:02}
{Miyaji}, T.,  \& {Griffiths}, R.~E. 2002, \apjl, 564, L5

\bibitem[\protect\citeauthoryear{{Moore} et~al.}{{Moore}
  et~al.}{1999}]{Moore:99}
{Moore}, B., {Ghigna}, S., {Governato}, F., {Lake}, G., {Quinn}, T., {Stadel},
  J.,  \& {Tozzi}, P. 1999, \apjl, 524, L19

\bibitem[\protect\citeauthoryear{{Oh}}{{Oh}}{2001}]{Oh:00}
{Oh}, S.~P. 2001, \apj, 553, 499

\bibitem[\protect\citeauthoryear{{Oh} \& {Haiman}}{{Oh} \&
  {Haiman}}{2002}]{OhH:02}
{Oh}, S.~P.,  \& {Haiman}, Z. 2002, \apj, 569, 558

\bibitem[\protect\citeauthoryear{{Pasquali} et~al.}{{Pasquali}
  et~al.}{2005}]{Pasquali:05}
{Pasquali}, A., {Larsen}, S., {Ferreras}, I., {Gnedin}, O.~Y., {Malhotra}, S.,
  {Rhoads}, J.~E., {Pirzkal}, N.,  \& {Walsh}, J.~R. 2005, \aj, 129, 148

\bibitem[\protect\citeauthoryear{{Pettini} et~al.}{{Pettini}
  et~al.}{2003}]{Pettini:03}
{Pettini}, M., {Madau}, P., {Bolte}, M., {Prochaska}, J.~X., {Ellison}, S.~L.,
  \& {Fan}, X. 2003, \apj, 594, 695

\bibitem[\protect\citeauthoryear{{Ricotti}}{{Ricotti}}{2003}]{Ricotti:03}
{Ricotti}, M. 2003, \mnras, 344, 1237

\bibitem[\protect\citeauthoryear{{Ricotti}, {Ferrara}, \& {Miniati}}{{Ricotti}
  et~al.}{1997}]{Ricotti:97}
{Ricotti}, M., {Ferrara}, A.,  \& {Miniati}, F. 1997, \apj, 485, 254

\bibitem[\protect\citeauthoryear{{Ricotti} \& {Gnedin}}{{Ricotti} \&
  {Gnedin}}{2005}]{RicottiG:05}
{Ricotti}, M.,  \& {Gnedin}, N.~Y. 2005, \apj, 629, 259

\bibitem[\protect\citeauthoryear{{Ricotti}, {Gnedin}, \& {Shull}}{{Ricotti}
  et~al.}{2001}]{RicottiGS:01}
{Ricotti}, M., {Gnedin}, N.~Y.,  \& {Shull}, J.~M. 2001, \apj, 560, 580

\bibitem[\protect\citeauthoryear{{Ricotti}, {Gnedin}, \& {Shull}}{{Ricotti}
  et~al.}{2002a}]{RicottiGSa:02}
{Ricotti}, M., {Gnedin}, N.~Y.,  \& {Shull}, J.~M. 2002a, \apj, 575, 33

\bibitem[\protect\citeauthoryear{{Ricotti}, {Gnedin}, \& {Shull}}{{Ricotti}
  et~al.}{2002b}]{RicottiGSb:02}
{Ricotti}, M., {Gnedin}, N.~Y.,  \& {Shull}, J.~M. 2002b, \apj, 575, 49

\bibitem[\protect\citeauthoryear{{Ricotti} \& {Ostriker}}{{Ricotti} \&
  {Ostriker}}{2004a}]{RicottiOI:03}
{Ricotti}, M.,  \& {Ostriker}, J.~P. 2004a, \mnras, 350, 539

\bibitem[\protect\citeauthoryear{{Ricotti} \& {Ostriker}}{{Ricotti} \&
  {Ostriker}}{2004b}]{RicottiO:03}
{Ricotti}, M.,  \& {Ostriker}, J.~P. 2004b, \mnras, 352, 547

\bibitem[\protect\citeauthoryear{{Ricotti}, {Ostriker}, \& {Gnedin}}{{Ricotti}
  et~al.}{2005}]{RicottiOG:03}
{Ricotti}, M., {Ostriker}, J.~P.,  \& {Gnedin}, N.~Y. 2005, \mnras, 357, 207

\bibitem[\protect\citeauthoryear{{Ricotti} \& {Wilkinson}}{{Ricotti} \&
  {Wilkinson}}{2004}]{RicottiW:03}
{Ricotti}, M.,  \& {Wilkinson}, M.~I. 2004, \mnras, 353, 867

\bibitem[\protect\citeauthoryear{{Santoro} \& {Shull}}{{Santoro} \&
  {Shull}}{2006}]{SantoroShull:06}
{Santoro}, F.,  \& {Shull}, J.~M. 2006, \apj, 643, 26

\bibitem[\protect\citeauthoryear{{Sasaki}}{{Sasaki}}{1994}]{Sasaki:94}
{Sasaki}, S. 1994, \pasj, 46, 427

\bibitem[\protect\citeauthoryear{{Schaye} et~al.}{{Schaye}
  et~al.}{2003}]{Schaye:03}
{Schaye}, J., {Aguirre}, A., {Kim}, T., {Theuns}, T., {Rauch}, M.,  \&
  {Sargent}, W.~L.~W. 2003, \apj, 596, 768

\bibitem[\protect\citeauthoryear{{Shull} \& {Venkatesan}}{{Shull} \&
  {Venkatesan}}{2008}]{ShullVenkatesan:07}
{Shull}, J.~M.,  \& {Venkatesan}, A. 2008, ApJ, in press

\bibitem[\protect\citeauthoryear{{Simcoe}, {Sargent}, \& {Rauch}}{{Simcoe}
  et~al.}{2004}]{Simcoe:04}
{Simcoe}, R.~A., {Sargent}, W.~L.~W.,  \& {Rauch}, M. 2004, \apj, 606, 92

\bibitem[\protect\citeauthoryear{{Simon} \& {Geha}}{{Simon} \&
  {Geha}}{2007}]{SimonG:07}
{Simon}, J.~D.,  \& {Geha}, M. 2007, \apj, 670, 313

\bibitem[\protect\citeauthoryear{{Songaila}}{{Songaila}}{2001}]{Songaila:01}
{Songaila}, A. 2001, \apjl, 561, L153

\bibitem[\protect\citeauthoryear{{Spergel} et~al.}{{Spergel}
  et~al.}{2007}]{Spergel:07}
{Spergel}, D.~N., et~al. 2007, \apjs, 170, 377

\bibitem[\protect\citeauthoryear{{Susa} \& {Umemura}}{{Susa} \&
  {Umemura}}{2004}]{Susa:04}
{Susa}, H.,  \& {Umemura}, M. 2004, \apj, 600, 1

\bibitem[\protect\citeauthoryear{{Tassis} et~al.}{{Tassis}
  et~al.}{2003}]{Tassis:03}
{Tassis}, K., {Abel}, T., {Bryan}, G.~L.,  \& {Norman}, M.~L. 2003, \apj, 587,
  13

\bibitem[\protect\citeauthoryear{{Tegmark} et~al.}{{Tegmark}
  et~al.}{1997}]{Tegmark:97}
{Tegmark}, M., {Silk}, J., {Rees}, M.~J., {Blanchard}, A., {Abel}, T.,  \&
  {Palla}, F. 1997, \apj, 474, 1

\bibitem[\protect\citeauthoryear{{Tumlinson}}{{Tumlinson}}{2006}]{Tumlinson:06}
{Tumlinson}, J. 2006, \apj, 641, 1

\bibitem[\protect\citeauthoryear{{Venkatesan}, {Giroux}, \&
  {Shull}}{{Venkatesan} et~al.}{2001}]{Venkatesan:01}
{Venkatesan}, A., {Giroux}, M.~L.,  \& {Shull}, J.~M. 2001, \apj, 563, 1

\bibitem[\protect\citeauthoryear{{Venkatesan} \& {Truran}}{{Venkatesan} \&
  {Truran}}{2003}]{VenkatesanT:03}
{Venkatesan}, A.,  \& {Truran}, J.~W. 2003, \apjl, 594, L1

\bibitem[\protect\citeauthoryear{{Walsh}, {Jerjen}, \& {Willman}}{{Walsh}
  et~al.}{2007}]{Walshetal07}
{Walsh}, S.~M., {Jerjen}, H.,  \& {Willman}, B. 2007, \apjl, 662, L83

\bibitem[\protect\citeauthoryear{{Whalen} \& {Norman}}{{Whalen} \&
  {Norman}}{2008}]{Whalen:08}
{Whalen}, D.,  \& {Norman}, M.~L. 2008, \apj, 673, 664

\bibitem[\protect\citeauthoryear{{Whiting}, {Hau}, \& {Irwin}}{{Whiting}
  et~al.}{1999}]{Whiting:99}
{Whiting}, A.~B., {Hau}, G.~K.~T.,  \& {Irwin}, M. 1999, \aj, 118, 2767

\bibitem[\protect\citeauthoryear{{Willman} et~al.}{{Willman}
  et~al.}{2005a}]{Willmanetal05AJ}
{Willman}, B., et~al. 2005a, \aj, 129, 2692

\bibitem[\protect\citeauthoryear{{Willman} et~al.}{{Willman}
  et~al.}{2005b}]{Willmanetal05ApJ}
{Willman}, B., et~al. 2005b, \apjl, 626, L85

\bibitem[\protect\citeauthoryear{{Wolfire} et~al.}{{Wolfire}
  et~al.}{1995}]{Wolfire:95}
{Wolfire}, M.~G., {Hollenbach}, D., {McKee}, C.~F., {Tielens}, A.~G.~G.~M.,  \&
  {Bakes}, E.~L.~O. 1995, \apj, 443, 152

\bibitem[\protect\citeauthoryear{{Wolfire} et~al.}{{Wolfire}
  et~al.}{2003}]{Wolfire:03}
{Wolfire}, M.~G., {McKee}, C.~F., {Hollenbach}, D.,  \& {Tielens}, A.~G.~G.~M.
  2003, \apj, 587, 278

\bibitem[\protect\citeauthoryear{{Yoshida} et~al.}{{Yoshida}
  et~al.}{2003}]{Yoshida:03}
{Yoshida}, N., {Abel}, T., {Hernquist}, L.,  \& {Sugiyama}, N. 2003, \apj, 592,
  645

\bibitem[\protect\citeauthoryear{{Zucker} et~al.}{{Zucker}
  et~al.}{2006a}]{Zuckeretal06b}
{Zucker}, D.~B., et~al. 2006a, \apjl, 650, L41

\bibitem[\protect\citeauthoryear{{Zucker} et~al.}{{Zucker}
  et~al.}{2006b}]{Zuckeretal06a}
{Zucker}, D.~B., et~al. 2006b, \apjl, 643, L103

\end{thebibliography}

\vskip 2truecm

\placefig{\end{document}}

\clearpage

\newcounter{figurecap}
\setcounter{figurecap}{0}

\begin{center}
\bf Figure Captions
\end{center}
\refstepcounter{figurecap}
Fig.\ \thefigurecap---\label{fig:tcool}{\capfigaa}

\refstepcounter{figurecap}
Fig.\ \thefigurecap---\label{fig:bias1}\capfigba

\refstepcounter{figurecap}
Fig.\ \thefigurecap---\label{fig:bias2}\capfigca

\refstepcounter{figurecap}
Fig.\ \thefigurecap---\label{fig:fill2}\capfigda

\refstepcounter{figurecap}
Fig.\ \thefigurecap---\label{fig:bar1}\capfigdb

\refstepcounter{figurecap}
Fig.\ \thefigurecap---\label{fig:starf1}\capfigea

\refstepcounter{figurecap}
Fig.\ \thefigurecap---\label{fig:starf2}\capfigeb

\refstepcounter{figurecap}
Fig.\ \thefigurecap---\label{fig:zmasf1}\capfigfa

\refstepcounter{figurecap}
Fig.\ \thefigurecap---\label{fig:nick}\capfigga

\refstepcounter{figurecap}
Fig.\ \thefigurecap---\label{fig:prof}\capfiggb

\refstepcounter{figurecap}
Fig.\ \thefigurecap---\label{fig:phase}\capfiggc

\refstepcounter{figurecap}
Fig.\ \thefigurecap---\label{fig:NGST}\capfigha

\refstepcounter{figurecap}
Fig.\ \thefigurecap---\label{fig:Ncount}\capfighb

\refstepcounter{figurecap}
Fig.\ \thefigurecap---\label{fig:massf}\capfighc

\clearpage

\tabone

\tabtwo

\end{document}